\documentclass[preprint,nofootinbib]{revtex4}%
\usepackage[T1]{fontenc}
\usepackage{amsmath}
\usepackage{amssymb}
\usepackage{esint}
\usepackage{amsfonts}
\usepackage{graphicx,color}
\usepackage{slashed}
\usepackage{young}
\usepackage[vcentermath]{youngtab}
\usepackage[utf8]{inputenc}
\usepackage{tensor}
\usepackage{etoolbox}
\usepackage{bbm}
\usepackage{soul}
\usepackage{ulem}

\usepackage{epic,eepic,pstricks}
\usepackage{mathrsfs}

\makeatletter

\newcommand{\be}{\begin{equation}}
\newcommand{\ee}{\end{equation}}
\newcommand{\bea}{\begin{eqnarray}}
\newcommand{\eea}{\end{eqnarray}}

\renewcommand{\tilde}{\widetilde}
\renewcommand{\hat}{\widehat}

\renewcommand{\d}{\partial}

\@addtoreset{equation}{section}

\newcommand{\RR}{\mathbb{R}}

\newcommand{\NN}{\mathbb{N}}

\def\cL{\mathcal{L}}

\newcommand*\xbar[1]{%
  \hbox{%
    \vbox{%
      \hrule height 0.5pt 
      \kern0.3ex
      \hbox{%
        \kern-0.0em
        \ensuremath{#1}%
        \kern-0.0em
      }%
    }%
  }%
} 

\pdfpageheight\paperheight
\pdfpagewidth\paperwidth

\pdfoutput=1 

    \patchcmd{\maketitle}{\@fpheader}{}{}{}

\begin{document}

\title{The asymptotic structure of electromagnetism in higher spacetime dimensions}
\author{Marc Henneaux${}^1$\footnote{On leave of absence from Coll\`ege de France, 11 place Marcelin Berthelot, 75005 Paris, France} and C\'edric Troessaert${}^2$}
\affiliation{${}^1$Universit\'e Libre de Bruxelles and International Solvay Institutes, ULB-Campus Plaine CP231, B-1050 Brussels, Belgium}
\affiliation{${}^2$Max-Planck-Institut f\"{u}r Gravitationsphysik (Albert-Einstein-Institut),
Am M\"{u}hlenberg 1, \\ DE-14476 Potsdam, Germany}

\begin{abstract}
{We investigate the asymptotic structure of electromagnetism in Minkowski space in  even and odd spacetime dimensions $\geq 4$. We focus on $d>4$ since the case $d=4$ has been studied previously at length. We first consider spatial infinity where we provide explicit boundary conditions  that admit the known physical solutions and make the formalism well defined (finite symplectic structure and charges).   Contrary to the situation found in $d=4$ dimensions, there is no need to impose parity conditions under the antipodal map on the leading order of the fields when $d>4$. There is, however, the same need to modify the standard bulk symplectic form by a boundary term at infinity involving  a surface degree of freedom.  This step makes  the Lorentz boosts act canonically.   Because of the absence of parity conditions, the theory is found to be invariant under two independent algebras of angle-dependent $u(1)$ transformations ($d>4$). 
We then integrate the equations of motion in order to find the behaviour of the fields near null infinity.  We exhibit the radiative and Coulomb branches, characterized by different decays and parities.   The analysis yields generalized matching conditions between the past of $\mathscr{I}^+$ and the future of $\mathscr{I} ^-$. }
\end{abstract}



\maketitle

\section{Introduction}
\setcounter{equation}{0}

Most of the studies of the asymptotic properties of gravity in the asymptotically flat context have been performed at null infinity \cite{Bondi:1962px,Sachs:1962wk,Sachs:1962zza,Penrose:1962ij} (for recent useful reviews, see \cite{Madler:2016xju,Alessio:2017lps,Ashtekar:2018lor}).  This is quite natural, and would seem to be even mandatory, in order to decipher the intricate properties of gravitational radiation.  

One conceptual difficulty with analyses at null infinity, however, is that the existence of a null infinity with the smoothness properties usually assumed in the asymptotic treatments is a difficult dynamical question: given reasonable initial data on a Cauchy hypersurface, will their Cauchy development give rise to a null infinity with the requested properties?  Strong doubts that this would be the case  have been expressed in  \cite{Christodoulou:1993uv}, which we quote verbatim:

``... {\it it remains questionable whether there exists any non-trivial solution of
the field equations that satisfies the Penrose requirements} [of asymptotic simplicity].{\it Indeed, his regularity assumptions
translate into fall-off conditions of the curvature that may be too stringent and thus may fail
to be satisfied by any solution that would allow gravitational waves.}'' \\
This point has also been forcefully  stressed in the recent work \cite{Friedrich:2017cjg}.

One remarkable by-product of the studies at null infinity was the discovery that the asymptotic symmetry group of gravity in the asymptotically flat context was the infinite-dimensional Bondi-Metzner-Sachs (BMS) group.  Initially received with some skepticism because the physical significance of this infinite-dimensional enlargement of the Poincar\'e group was not clear, the emergence of the BMS group was understood recently to be related to profound infrared properties of gravity having to do with soft graviton theorems and memory effects  \cite{Strominger:2013jfa,He:2014laa,Cachazo:2014fwa,Strominger:2014pwa,Pasterski:2015tva,Campiglia:2015kxa,Conde:2016rom}
(see \cite{Strominger:2017zoo} for an exposition of this recent work and \cite{Ashtekar:1981bq,Ashtekar:1981sf,Ashtekar:1987tt} for earlier investigations).  The conclusion of the huge amount of activity that florished since then is that the BMS group is a gift rather than an embarrassment!  An even further enlargement of the Poincar\'e group including ``super-rotations'' have been even argued to be useful \cite{Banks:2003vp,Barnich:2010eb,Barnich:2009se}.

The BMS transformations are diffeomorphisms leaving the boundary conditions at null infinity invariant. They are exact symmetries of the theory.  That is, they leave the action exactly invariant up to a surface term, without having to make approximations.  As exact symmetries of the theory, they should be visible in any description, and, in particular, in slicings of spacetime adapted to spatial infinity.  In such slicings, they would appear as diffeomorphisms leaving the boundary conditions at spatial infinity invariant.  For this to be the case, however, the boundary conditions at spatial infinity should be equivalent, or at least compatible in a sense that we shall make more precise below, with the boundary conditions at null infinity.  This brings us back to the dynamical question on null infinity mentioned above.

Earlier investigations of the asymptotic symmetries at spatial infinity showed no sign of the BMS group.  One either found the Poincar\'e group with no enlargement \cite{Arnowitt:1962hi,Regge:1974zd}, or the smaller homogeneous Lorentz group \cite{Geroch:1972up}, or an even larger extension, the Spi group \cite{Ashtekar:1978zz,Ashtekar:1991vb}, but in no case the BMS group uncovered at null infinity.  One logical possibility for this descrepancy would be that the boundary conditions at spatial infinity are incompatible with the boundary conditions at null infinity, so that the set of transformations preserving ones would not preserve the others\footnote{Invariance of the action  cannot be the issue - provided the action is well-defined - since we are dealing with diffeomorphisms.}.   

If true, this situation would be very disappointing and physically unsatisfactory. Motivated by the desire to understand better these earlier puzzling results, we have re-examined the asymptotic structure of gravity at spatial infinity \cite{Henneaux:2018hdj,Henneaux:2018cst}.  We have provided in \cite{Henneaux:2018hdj} boundary conditions at spatial infinity that eliminate the previous tensions between spatial infinity and null infinity analyses, in the sense that: (i) these boundary conditions are invariant under the BMS group, which acts non trivially on the fields and has generically non vanishing conserved charges; (ii) integration of the symmetry generators from spatial to null infinity enables one to show that it is the same BMS group that acts both at spatial infinity and at null infinity, expressed in different parametrizations that can be explicitly related \cite{Troessaert:2017jcm}.  

Furthermore, the matching conditions imposed at null infinity on the leading order of the gravitational field \cite{Strominger:2017zoo} are automatic consequences of the asymptotic behaviour of the Cauchy data at spatial infinity\footnote{Although not equivalent (they are stronger), the boundary conditions at null infinity are compatible with those at spatial infinity, in the sense that they obey the conditions  that are implied at null infinity by the behaviour at spatial infinity.}.  It is of interest to point out in this respect that while the leading order of the Cauchy development of the gravitational field coincides with the generally assumed leading order at null infinity, the subsequent terms in the expansion differ in general, since subleading terms of the type $\frac{\ln r}{r^k}$ ($ k \geq 1$) will develop from generic initial data.  One consequence of our analysis is that these non-analytic terms do not spoil the BMS symmetry -- even if they spoil the usually assumed ``peeling'' behaviour of the gravitational field at null infinity \cite{Friedrich:2017cjg}.  This gives further robustness to the BMS symmetry\footnote{Incidentally, in our first work on this problem \cite{Henneaux:2018cst}, we put forward alternative boundary conditions that were also BMS invariant, but which yielded a singular behaviour ($\sim \ln r$) for some components of the Weyl tensor as one went to null infinity.  In spite of this singular behaviour at null infinity, nothing spectacular occurred at spatial infinity and the BMS symmetry was untouched.}.  It also disentangles the BMS group from gravitational radiation.

Similar features arise in the discussion of the asymptotic behaviour of the electromagnetic field, where the null infinity analysis \cite{Strominger:2013lka,Barnich:2013sxa,He:2014cra} seemed to be at variance with the spatial infinity analysis \cite{Henneaux:1999ct}.  The tension was solved in \cite{Henneaux:2018gfi}, again by providing appropriate boundary conditions at spatial infinity.  The null infinity matching conditions of electromagnetism were also shown there to be a consequence of the boundary conditions at spatial infinity. 

Extension of the asymptotic analysis to higher dimensions raises interesting issues, which have led to a somewhat unclear situation at null infinity where some studies yield infinite-dimensional asymptotic symmetries as in four spacetime dimensions, while some others do not \cite{Tanabe:2012fg,Kapec:2015vwa,Hollands:2016oma,Garfinkle:2017fre,Mao:2017wvx,Campiglia:2017xkp,Pate:2017fgt,Campoleoni:2017qot,Afshar:2018apx,Campoleoni:2018uib}.  The question is further complicated in odd spacetime dimensions because half-integer fractional powers of $r^{-1}$ mix with integer powers, leading to problems with the conformal definition of null infinity \cite{Hollands:2003ie,Hollands:2004ac,Tanabe:2011es}, and the frequent necessity to split the analysis according to whether the spacetime dimension is odd or even since only in the latter case does one avoid non-analytic functions at null infinity.

This provides strong motivations for investigating the asymptotic structure of the electromagnetic and gravitational fields at spatial infinity in higher dimensions, where the fall-off of the fields is more uniform (no fractional powers of $r$). This is done here for electromagnetism.  We show that the methods developed in our previous work  \cite{Henneaux:2018gfi} generalize straightforwardly to higher dimensions, with no new conceptual  difficulty.   The discussion proceeds along similar lines independently of the spacetime dimension.  One finds in particular the same need to modify the standard bulk symplectic structure by a surface term, as shown necessary also by different methods in $d=4$ spacetime dimensions \cite{Campiglia:2017mua,Campiglia:2018dyi}.

One remarkable feature, however, is that a second angle-dependent $u(1)$ asymptotic symmetry emerges.  This second $u(1)$ is eliminated in $4$ spacetime dimensions because of parity conditions that must be imposed to get rid of divergences in the symplectic structure and of divergences in some components of the fields as one goes to null infinity \cite{Henneaux:2018gfi}, but these parity conditions turn out to be unnecessary in higher dimensions (although it would be consistent to impose them).  

The difference in the behaviour of the fields according to whether the dimension is even or odd appears when one considers null infinity.  We exhibit the behaviour of the electromagnetic field near null infinity by integrating the equations of motion ``from spatial infinity to null infinity''.  This is done by going first to hyperbolic coordinates \cite{Ashtekar:1978zz,BeigSchmidt,Beig:1983sw,Compere:2011ve}.  Hyperbolic coordinates are pathological in the limit, however, and we thus go then to coordinates introduced by 
Friedrich, which are better suited to that purpose \cite{Fried1,Friedrich:1999wk,Friedrich:1999ax}.  We show that initial data fulfilling our asymptotic conditions at spatial infinity, without parity conditions, lead to a non-divergent behaviour at null infinity ($d>4$).  The presence of terms with different parities leads to an interesting generalization of the matching conditions between fields at the past of $\mathscr{I}^+$ and the future of $\mathscr{I}^-$, which we give.

Our paper is organized as follows.  Section \ref{subsec:Starting} provides the boundary conditions for the standard canonical variables for (free) electromagnetism, i.e., the spatial components of the vector potential and their conjugate momenta, which are the components of the electric field. We focus on the case of $d=5$ spacetime dimensions. The symplectic form is finite without parity conditions.  We allow a gradient term  $\d_i \Phi$ where $\Phi$  is of order $O(r^0)$ in the asymptotic behaviour of the vector potential.  This gives a gauge invariant formulation of the boundary conditions, and is crucial for exhibiting the full set of asymptotic symmetries.  In Section \ref{sec:ProperImproper}, we determine the proper and improper \cite{Benguria:1976in} asymptotic symmetries for these boundary conditions.  We show in Section \ref{sec:PoincareTransf} that because of the presence of a gradient term in the boundary conditions, the Lorentz boosts are not canonical transformations.  The problem can be cured by introducing a surface degree of freedom (which can ultimately be identified with $A_0$ at the boundary). This is just as in $4$ spacetime dimensions \cite{Henneaux:2018gfi}.  We then give the complete formulation of the $d=5$ theory in Section \ref{sec:55}  where we write in particular explicitly all the Poincar\'e generators.  In Section \ref{sec:AsympSym2}, we  show  how the second angle-dependent $u(1)$ symmetry emerges.  Section \ref{sec:More5} generalizes the analysis to arbitrary spacetime dimension $\geq 5$. The detailed behaviour of the fields as one goes to null infinity  is derived in Section \ref{sec:NullInf}, where we compare and contrast the situations in $d=4$ spacetime dimensions (where parity conditions are necessary to remove leading logarithmic divergences) and $d>4$ spacetime dimensions (where this is not necessary).  The concluding Section \ref{sec:Conclusions} gives further light on the emergence of the second angle-dependent $u(1)$.  Three appendices complete the discussion.

As it is common practice in such asymptotic investigations, we shall
assume ``uniform smoothness'' \cite{Sachs:1962zza} whenever needed, i.e., $\d_r o(r^{-k}) =
o(r^{-k-1}), \d_A o(r^{-k}) = o(r^{-k})$.  Similarly, the distinction between $O(r^{-(k+1)})$ and $o(r^{-k})$ will usually not be important to the orders relevant to the analysis.

We close this introduction by recalling what is meant here by the concept of asymptotic symmetry.  This concept is defined only in space with boundaries (which can be infinity), once boundary conditions are prescribed to complete the definition of the theory \cite{Fock}, as particularly emphasized in \cite{Deser:2019acl}. Asymptotic symmetries are gauge transformations that preserve the boundary conditions (and yield  finite surface terms in the variation of the action so as to have well-defined canonical generators, here at spatial infinity).  An asymptotic symmetry is non trivial if its generator is not identically zero, i.e., if there are allowed configurations (configurations obeying the boundary conditions) that make it not vanish.  Such an asymptotic symmetry is then called ``improper gauge symmetry'' following the terminology introduced in the lucid paper \cite{Benguria:1976in}.  ``Proper gauge transformations'' have identically vanishing generators and form an ideal.  The true physical asymptotic symmetry algebra is the quotient of all the asymptotic symmetries by the proper ones.  Note that no gauge condition is involved in that definition, which is therefore intrinsic since it does not view asymptotic symmetries as residual gauge transformations preserving some gauge conditions\footnote{Of course, the boundary conditions might involve implicitly some gauge fixing as it is usually difficult to formulate them in terms of gauge invariants only.  It is important to check gauge independence.  The boundary conditions given below for electromagnetism leave the freedom of making an arbitrary gauge transformation with finite generator.}.

\section{Action and boundary conditions -- Preliminary considerations}
\label{subsec:Starting}
\setcounter{equation}{0}

We start with the standard action of source-free electromagnetism in $d$ spacetime dimensions, which takes the canonical form
\begin{equation}
\label{eq:StartingPoint}
	S_H[A_i, \pi^i, A_0] = \int dt \left\{ \int d^{d-1}x \, 
		\pi^i \d_t
		A_i 
		 	- \int d^{d-1}x \left(   
				\frac{1}{2} \pi^i \pi_i + \frac{1}{4} F^{ij} F_{ij}  + A_t  \mathcal{G}\right) + F_\infty\right\}
\end{equation}
where $F_\infty$ is a surface term at spatial infinity ($r \rightarrow \infty$),
which depends on the boundary conditions and which will be discussed below.
The dynamical variables to be varied in the action are the spatial components $A_i$ of the vector potential,  their conjugate momenta $\pi^i$ (equal to the electric field) and the temporal component $A_0 \equiv A_t$ of the vector potential which plays the role of Lagrange multiplier  for the constraint
\be
\mathcal{G} = - \partial_i \pi^i \approx 0
\ee
(Gauss' law).   We
use the symbol $\approx$ to denote equality on the constraint's surface. 

\subsection{Asymptotic behaviour of the fields: first conditions}

We now specialize to $4+1$ spacetime dimensions for
definiteness. In $4+1$ spacetime
dimensions, the electric and magnetic fields decay at spatial infinity as
$\frac{1}{r^{3}}$.  This implies that the electromagnetic potential
behaves as $\frac{1}{r^2}$ up to a gauge transformation, i.e., up to a
gradient $\partial_i \Phi$.  Under a gauge transformation $A_i \rightarrow
A_i + \partial_i \epsilon$, $\Phi$ is shifted by $\epsilon$.  This transformation will have a well-defined generator if $\epsilon$ is of order unity at infinity (see below).  It is therefore natural to request that $\Phi$ be also of order unity at infinity  (which implies $\partial_i \Phi \sim \frac{1}{r}$).  We thus impose the following decay at  spatial infinity,
\begin{equation}
	A_i = \partial_i \xbar \Phi + \frac{1}{r^2} \xbar A_i + \frac{1}{r^3} A^{(1)}_i +
	o(r^{-3}), \quad 
	\pi^i = \frac{1}{r^3} \xbar \pi^i + \frac{1}{r^4} \pi^{(1)i} +
	o(r^{-4}) \label{GenFallOff}
\end{equation}
where $\xbar \Phi$ and the coefficients of the various powers of $r^{-1}$ are arbitrary functions on the $3$-sphere, i.e., of the angles $x^A$ used to parametrize it.  We have kept only the leading term in $\Phi = \xbar \Phi + \frac{\Phi^{(1)}}{r} + O(\frac{1}{r^2})$, since the subsequent terms can be absorbed by redefinitions.  It is only in $3+1$ dimensions that $\partial_i \xbar \Phi$ and the first term containing $\xbar A_i$ are of the same order.  Here, $\partial_i \xbar \Phi$ decays more slowly by one power of $r$.

The boundary conditions (\ref{GenFallOff}) make the kinetic term in the action well-defined provided we impose that Gauss'law holds at infinity one order faster than expected, i.e.
\begin{equation}
\partial_i \pi^i = o(\frac{1}{r^{4}})
\label{eq:FastDecayGauss}
\end{equation}
(the order implied by (\ref{GenFallOff}) is $o(\frac{1}{r^{3}})$).  This ensures that the term $\int d^4x \, \pi^i \d_t \partial_i  \Phi $ has no logarithmic singularity and clearly does not eliminate any physical solution.  The condition (\ref{eq:FastDecayGauss}) is an integral part  of our boundary conditions.

The asymptotic conditions (\ref{eq:FastDecayGauss}) are sufficient by themselves to make the symplectic form finite.  There is no need to impose parity conditions on the fields, contrary to what was found in 4 spacetime dimensions \cite{Henneaux:2018gfi}, where an appropriate generalization of the parity conditions of \cite{Regge:1974zd,Henneaux:1999ct} was necessary.

\subsection{Polar coordinates}

For later purposes, we rewrite the boundary conditions in spherical coordinates, in which the Minkowski metric reads
\be
ds^2 = - dt^2 + dr^2 + g_{AB} dx^A dx^B,
\ee with
\be
g_{AB} = r^2 \xbar \gamma_{AB}
\ee
where $\gamma_{AB}$ is the round metric in the unit sphere.  In the sequel angular indices on a ``bar'' quantity will be raised or lowered with $\xbar \gamma_{AB}$, e.g., $\xbar v^A = \xbar \gamma^{AB} \xbar v_B$.  Bar quantities live on the unit sphere. 

 One gets for the asymptotic fall-off in polar coordinates, recalling that the momenta carry a unit density weight:
\begin{gather}
	\label{eq:hamilasymptI0}
	A_r = \frac{1}{r^2} \xbar A_r + \frac{1}{r^3} A^{(1)}_r +
	o(r^{-3}),\quad
	\pi^r = \xbar \pi^r + \frac{1}{r} \pi^{(1)r} +
	o(r^{-1}),\\
	A_A = \partial_A \xbar \Phi + \frac{1}{r}\xbar A_A + \frac{1}{r^2} A^{(1)}_A +
	o(r^{-2}),\quad
	\pi^A = \frac{1}{r}\xbar \pi^A + \frac{1}{r^2} \pi^{(1)A} +
	o(r^{-2}), \label{eq:hamilasymptI022}
\end{gather}
where the coefficients of the various powers of $1/r$ are functions of the angles $x^A$. Furthermore,
\begin{equation}
\partial_A \xbar \pi^A = 0.
\label{eq:hamilasymptI036}
\ee
Note that $\partial_r \Phi$ is of order $O(\frac{1}{r^2})$ and has been absorbed in $\xbar A_r$.

\subsection{Relativistic invariance of the boundary conditions}

The boundary conditions (\ref{GenFallOff}), (\ref{eq:FastDecayGauss}) are easily verified to be Lorentz invariant.  A general deformation of a spacelike  hyperplane can be decomposed into normal and tangential components, denoted by $\xi$ and $\xi^i$, respectively. A Poincar\'e transformation corresponds to the deformation
\begin{eqnarray}
 \xi &=& b_i x^i + a^\perp  \label{eq:Poinc1}\\
 \xi^i &=&{ b^i}_j x^j + a^i  \label{eq:Poinc2}
\end{eqnarray}
where $b_i$, $b_{ij} = -b_{ji}$, $a^\perp$ and $a^i$ are arbitrary constants. The constants $b_i$ parametrize the Lorentz
boosts, whereas the antisymmetric constants $b_{ij} = -b_{ji}$
parametrize the spatial rotations.  The constants $a^\perp$ and $a^i$ are standard translations. 

Under such a deformation, the fields transform as
\begin{eqnarray}
&&\delta A_i = \xi \pi_i  + \xi^j F_{ji}  + \partial_i \zeta \label{eq:GTA}\\
&& \delta \pi^i = \partial_m \left( F^{mi} \xi \right) + \partial_m \left(\xi^m \pi^i \right) - (\partial_m \xi^i )\pi^m - \xi^i \partial_m \pi^m ,
\end{eqnarray}
The transformation of the fields is really defined up to a gauge transformation.  This is the reason why we have included the term $\partial_i \zeta$ in the transformation of $A_i$.  A definite choice  of accompanying gauge transformation will be made below to get simple expressions for the algebra.
It is clear that the fall-off (\ref{GenFallOff}), (\ref{eq:FastDecayGauss}) is preserved under these  transformations provided $\zeta$ behaves as $\zeta = \xbar \zeta(x^A) + \frac{\zeta^{1}(x^A)}{r} + o(\frac{1}{r^2})$.

The above transformation rules imply that the leading order of the fields  transform only under boosts and rotations (their variations under translations are of lower order).  One gets explicitly the following changes of the leading orders of the fields, which we write in polar coordinates,
\begin{gather}
	\delta_{b,Y} \xbar A_r = \frac{b}{\sqrt{\xbar \gamma}} \xbar \pi^r +
	Y^A \left( \d_A \xbar A_r + \xbar A_A \right) - \zeta^{1},  \label{eq:TransLead2}\\
	\delta_{b,Y} \xbar A_A = \frac{b}{\sqrt{\xbar\gamma}} \xbar
	\gamma_{AB}\xbar \pi^B + Y^B \left(\d_B \xbar A_A -   \d_A \xbar
	A_B\right) + \partial_A \zeta^{1}, \label{eq:TransLead2bis}\\
	\delta_{b,Y} \xbar \Phi = \xbar \zeta, \label{eq:TransLead2ter}\\
	\delta_{b,Y} \xbar \pi^r = \sqrt {\xbar \gamma}\, \xbar D^A\left(b (\d_A \xbar A_r+ \xbar A_A)\right) 
	+ \d_A(Y^A \xbar \pi^r), \label{eq:TransLead3}\\
	\delta_{b,Y} \xbar \pi^A = \sqrt {\xbar \gamma}\,\xbar D_B \Big(b\, \xbar \gamma^{BC}
	\xbar \gamma^{AD}(\d_C \xbar A_D - \d_D \xbar A_C)\Big) \nonumber \\  \hspace{1.5cm} + \d_B(Y^B \xbar \pi^A) - \d_B Y^A \xbar
	\pi^B - Y^A \partial_B \xbar \pi^B \label{eq:TransLead4}
\end{gather}
Here, we have set
\begin{gather}
	\xi = rb + T,\quad \xi^r = W, \quad \xi^A = Y^A + \frac 1 r \xbar D^A
	W,  \label{eq:FormOfW}\\
	\xbar D_A \xbar D_B W + \xbar \gamma_{AB} W = 0, \quad
	\xbar D_A \xbar D_B b + \xbar \gamma_{AB} b = 0, \quad \cL_Y \xbar
	\gamma_{AB} = 0, \quad \d_A T = 0.
\end{gather}
The quantities $b$, $Y^A$, $T$ and $W$ are functions on the sphere.  The first
two, $b$ and $Y^A$, describe the homogeneous Lorentz transformations, while
$T$ and $W$, which do not appear in the transformation laws
(\ref{eq:TransLead2})-(\ref{eq:TransLead4}) of the leading orders, describe the translations.  In these equations,  $\xbar D_A$ is the covariant derivative associated with 
$\xbar \gamma_{AB}$  and $\xbar D^A = \xbar  \gamma^{AB} \xbar D_B$.

Contrary to what happens in 4 spacetime dimensions, the leading order $\xbar A_r$ is not gauge invariant.  Furthermore, the boosts mix radial and angular components, which do not have independent dynamics anymore.  Finally, we note that the spacetime translations have no action on the leading orders of the fields.  The symmetry group at this order is thus the homogeneous Lorentz group.  To get the full Poincar\'e group, one needs to consider the subsequent terms in the asymptotic expansion of the fields\footnote{It is actually well known that in the case of gravity, the next-to-leading terms contribute crucially to the angular momentum and the boost charges and so cannot be ignored.}.

\subsection{Further strengthening of the boundary conditions}

It turns out that while the boosts preserve the boundary conditions, they fail to be canonical transformations (see Section \ref{sec:PoincareTransf} below).  In order to recover a canonical action for the boosts, one adds new surface degrees of freedom.  This can be achieved along the lines of \cite{Henneaux:2018gfi} if one  strengthen further the boundary conditions. 

We impose that the leading term $\xbar A_A$ of the angular part of the vector potential be a pure gradient. This condition is preserved under the Poincar\'e group if we also impose $\xbar \pi^A = 0$.  Thus, we complete (\ref{GenFallOff}), (\ref{eq:FastDecayGauss}) by
\be
\xbar A_A = \partial_A \xbar \Theta \, , \qquad \xbar \pi^A = 0. \label{eq:FAB=0}
\ee

The requirement $\xbar A_A = \partial_A \xbar \Theta$, which is new with respect to $d=4$ where it is not needed, is equivalent to the condition $\xbar F_{AB} = 0$.  Imposing $\xbar F_{AB} = 0$ does not seem unreasonable in 5 (and higher) spacetime dimensions.   If there are only electric sources, this condition is certainly fulfilled.  So the question is whether it eliminates interesting magnetic configurations.  In $4+1$ dimensions, magnetic sources are extended objects, namely, strings.  They can be of two different types: (i) infinite, or (ii) closed (strings with boundary do not yield a conservation law).  Our formalism only applies to the second case,  since we are not covering infinitely extended strings going all the way to infinity, for which there are extra degrees of freedom at infinity -- the endpoints of the string -- that must be taken into account (on which the Poincar\'e group acts).   But if the strings are closed, their field is more like the field of a dipole because the total charge that they carry is zero \cite{Teitelboim:1985yc}.  The magnetic field decays thus  faster at infinity (see Appendix \ref{app:MagnSources}) and it is legitimate to assume $\xbar F_{AB} = 0$. 

\subsection{Summary: complete set of boundary conditions at spatial infinity on $A_i$, $\pi^!$}
To summarize: the complete set of boundary conditions on the canonical variables $A_i$, $\pi^i$ is given by  (\ref{eq:hamilasymptI0})-(\ref{eq:hamilasymptI022}) supplemented by (\ref{eq:FAB=0}) (which automatically implies (\ref{eq:hamilasymptI036})).  We rewrite them here:
\begin{gather}
	A_r = \frac{1}{r^2} \xbar A_r + \frac{1}{r^3} A^{(1)}_r +
	o(r^{-3}),\quad
	\pi^r = \xbar \pi^r + \frac{1}{r} \pi^{(1)r} +
	o(r^{-1}),\\
	A_A = \partial_A \xbar \Phi + \frac{1}{r}\d_A\xbar \Theta + \frac{1}{r^2} A^{(1)}_A +
	o(r^{-2}),\quad
	\pi^A =  \frac{1}{r^2} \pi^{(1)A} +
	o(r^{-2}),\\
	\hbox{(Complete set of boundary conditions  on the canonical variables $A_i$, $\pi^i$)} \nonumber
\end{gather}
in order to have them conveniently grouped together.

Given the vector potential, the functions $\xbar \Phi (x^A)$ and $\xbar \Theta (x^A)$ are determined up to a constant.  We shall come back to this point later.

\section{Proper and improper gauge transformations}
\label{sec:ProperImproper}
\setcounter{equation}{0}

The boundary conditions (\ref{GenFallOff}) are invariant under gauge transformations generated by 
the first-class constraint-generator $\mathcal{G}$ :
\begin{equation}
	\delta_{\epsilon} A_i =  \d_i \epsilon, \quad
	\delta_{\epsilon} \pi^i = 0,  \label{eq:GaugeTrans0}
\end{equation}
provided the gauge parameter $\epsilon$ has the asymptotic behaviour
\begin{equation}
	\epsilon = \xbar \epsilon(x^A) + \frac{1}{r} \epsilon^{(1)}(x^A) +
	o(r^{-1}). \label{eq:GaugePar0}
\end{equation}

As already indicated above,  the leading term $\xbar A_r $ in the expansion of the radial component of the potential is not gauge invariant, contrary to what happens in $3+1$ spacetime dimensions.  It transforms as $\xbar A_r \rightarrow \xbar A_r - \epsilon^{(1)}$.   The leading  ($O(r^{-2}$)) term of the field strength $F_{Ar}$ is of course invariant.  It is given by $\xbar F_{Ar} = \partial_A \xbar A_r + \xbar A_A$.   One has $ \xbar A_A \rightarrow \xbar A_A + \partial_A \epsilon^{(1)}$ so that $\xbar F_{Ar} \rightarrow \xbar F_{Ar}$.

The generator of (\ref{eq:GaugeTrans0}) reads explicitly
\begin{eqnarray}
	G[\epsilon] &=& \int d^4x \, \epsilon \, \mathcal{G} + \oint d^3S_i \, \xbar \epsilon \, \xbar \pi^i  \approx
	\oint d^3S_i \, \xbar \epsilon \, \xbar \pi^i  \label{eq:ChargeG0} \\
	&\approx&
	\oint d^3x \, \xbar \epsilon \, \xbar \pi^r  \label{eq:ChargeG007}
\end{eqnarray}
(where the  integrations in the last terms are over the 3-sphere at infinity, i.e., over the angular variables $x^A$).

Another crucial difference with respect to $3+1$ spacetime dimensions is that since the leading order of the radial component $\pi^r$ does not fulfill particular parity conditions, both even and odd parts of the gauge parameters contribute to the charge  $G[\epsilon]$. Non-vanishing values of $\xbar \epsilon$ define ``improper gauge transformations'' \cite{Benguria:1976in}, no matter what the parity of $\xbar \epsilon \not= 0$ is.  An improper gauge transformation shifts the function $\xbar \Phi$ by $\xbar \epsilon$.

The value of the generator $G[\epsilon]$ for $\xbar \epsilon = \xbar \epsilon_0$ (constant) is the total electric charge $Q$. In the absence of charged matter field, $Q = G[\epsilon_0] = 0$, so that constant shifts of $\xbar \epsilon(x^A)$ are unobservable (no charged states). To have $Q \not=0$, one needs to couple charged sources to the electromagnetic field, in which case Gauss' law becomes $- \d_k \pi^k + j^0 \approx 0$. The asymptotic analysis is unchanged provided the sources are localized or decrease sufficiently rapidly at infinity.   We should stress, however, that even if $Q=0$, the generator $G[\epsilon]$ for non constant functions $\xbar \epsilon(x^A)$ on the sphere will generically be different from zero.

To complete the description of the asymptotic behaviour, we need to specify the fall-off of the Lagrange multiplier $A_t$.  Since   $A_t$ parametrizes the gauge transformation performed in the course of the evolution, we take for $A_t$ the same fall-off as for the gauge parameter $\epsilon$, 
\begin{equation}
	A_t = A_t^{(0)}(x^A)  + \frac{1}{r} \xbar A_t(x^A)+
	o(r^{-1}).
\end{equation}
 If $ A_t^{(0)}(x^A)  \not=0$,  the time evolution involves a non-trivial
improper gauge transformation.

\section{Boosts and symplectic structure}
\label{sec:PoincareTransf}
\setcounter{equation}{0}

We now turn to the question as to whether the Poincar\'e transformations, which preserve the boundary conditions as we have just seen, are canonical transformations. That is, we analyse whether they are true symmetries\footnote{Invariance of the symplectic form is a consequence of the invariance of the action up to a total time derivative.}.

We focus on boosts, which are the only transformations presenting difficulties.  For boosts, the above transformations reduce to
\begin{eqnarray}
&&\delta A_i = \xi \pi_i  + \partial_i \zeta \label{eq:GTABoosts}\\
&& \delta \pi^i = \partial_m \left( F^{mi} \xi \right)  
\end{eqnarray}
where $ \xi = br$. As we now show, these fail to be canonical transformations with the  symplectic $2$-form derived from (\ref{eq:StartingPoint}) 
\be
  \Omega = \int d^4x \, d_V \pi^i \, d_V A_ i
 \ee
 where the product is the exterior product $\wedge$ of forms which we are not writing explicitly, and where we use the symbol $d_V$ for the exterior derivative in phase space in order not to introduce confusion with the spacetime exterior derivative.

The transformation defined by the vector field $X$ is canonical if $d_V \left( i_X \Omega \right) = 0$.  Evaluating this expression for the boosts, one finds, by a computation that parallels the 3+1 case,
\be
d_V  \left( i_b \Omega \right) = \int d^4 x \, \d_m \left(\sqrt{g} \xi d_V F^{mi} \right) \, d_V A_i
\ee
where we have used $d_V \pi^i \, d_V \pi_i = 0$.  Integrating by parts and using $d_V F_{ij} \, d_V F^{ij} = 0$, we get that $d_V  \left( i_b \Omega \right)$ reduces to a surface term,
\be
d_V  \left( i_b \Omega \right) = \oint d^3 x \,  \sqrt{g} \, \xi \,  d_V F^{ri} \, d_V A_i \, ,
\ee
an expression that can be transformed to
\begin{equation}
d_V  \left( i_b \Omega \right) =  - \oint d^3 x  \sqrt{\xbar \gamma} \, b  \, d_V  \left(\xbar D ^A   \xbar A_r  +  \xbar A^A \right)\, d_V \partial_A \xbar \Phi 
\label{eq:unwanted}
\end{equation}
using the asymptotic form of the fields. 

This expression would vanish if we had not allowed a gradient term $\partial_i \Phi$ in $A_i$.  But with such a term, the variation of the symplectic form is generically non zero and  the boosts are accordingly non canonical transformations. 

In $3+1$ dimensions, the coefficient of the $ d_V \partial_A \xbar \Phi $-term reduces to $d_V \, \xbar D ^A   \xbar A_r $ because the contribution from the angular part is subleading.  This enables one to integrate by parts and replace the surface term by
\begin{equation}
d_V  \left( i_b \Omega \right) = \oint d^2 x  \sqrt{\xbar \gamma}  \, d_V  \xbar A_r  \,  \xbar D ^A (b \,  d_V \xbar  A_A)  \; \; \: \; (D = 3+1 \hbox{ dimensions})
\end{equation}
By introducing a single surface degree of freedom $\xbar \Psi$ that transforms appropriately under boosts 
and adding the surface contribution 
$
- \oint d^2x \, \sqrt{\xbar \gamma} \, d_V \xbar A_r \, d_V \xbar \Psi 
$
to the symplectic form, one can  make the Lorentz boosts canonical \cite{Henneaux:2018gfi}.

A similar route can be followed here provided $\xbar A_A = \partial_A \xbar \Theta$.  The variation of the bulk symplectic form  can then be transformed into
\begin{equation}
d_V  \left( i_b \Omega \right) = \oint d^3 x  \sqrt{\xbar \gamma}  \, d_V   \left(\xbar A_r + \xbar \Theta \right) \,  \xbar D ^A (b \,  d_V \partial_A \xbar  \Phi) 
\end{equation}
and by introducing a surface degree of freedom $\xbar \Psi$ at infinity transforming under boosts as
\be
\delta_b \xbar \Psi = \xbar D^A (b \partial_A \xbar \Phi)   \label{eq:VarPsi}
\ee
 and adding the surface term 
\be
- \oint d^3x \, \sqrt{\xbar \gamma} \, d_V \left(\xbar A_r +  \xbar \Theta \right)\, d_V \xbar \Psi \label{eq:SympSurfTerm}
\ee
one finds that the boosts are canonical (see detailed computation in Subsection \ref{sec:PoinCh1} below). 

The discussion proceeds in fact as in the 3+1 case, with the $3+1$ gauge-invariant $\xbar A_r$ replaced by the $4+1$ gauge-invariant $\xbar A_r + \xbar \Theta$.    The development parallels exactly the discussion in $4$ dimensions and is given in the next section.

There is, however, one important difference with respect to the $3+1$ dimensional case: it is that both even and odd parts of the the field $\xbar \Psi$ carry physical degrees of freedom. This is because the coefficient $\xbar A_r +  \xbar \Theta $ of $d_V \xbar \Psi $ in (\ref{eq:SympSurfTerm}) has no particular parity property (while it is odd in $3+1$ dimensions).

\section{Complete formulation}
\label{sec:55}
\setcounter{equation}{0}

We shall thus give only here the salient features. 
At this stage, the field $\xbar \Psi$ is  a field living on the three-sphere at infinity, which can depend on time.   As in $3+1$ dimensions, one can extend it inside the bulk to a ``normal'' field with conjugate momentum $\pi_\Psi$ constrained to vanish.  It is to this formulation that we shall immediately proceed.

\subsection{Action}

The complete action is
\begin{multline}
	S_H[A_i, \pi^i, \Psi, \pi_\Psi, \xbar \Theta_0; A_t, \lambda] = \int dt \left\{ \int d^4x \, 
		\pi^i \d_t
		A_i + \pi_\Psi \d_t \Psi - \oint d^3x
		\, \sqrt {\xbar\gamma}\,  \left(\xbar A_r +  \xbar \Theta \right) \, \d_t \xbar \Psi
		\right.\\
		\left. 	- \int d^4x \left(\frac{1}{2\sqrt g} \pi^i \pi_i + \frac{\sqrt
		g}{4} F^{ij} F_{ij} \right)  - \int d^4x \left( \lambda \pi_\Psi +  A_t \mathcal{G}  \right)\right\}.
\end{multline}
where $\Psi$ and $\pi_\Psi$ are new fields which behave at infinity as
\begin{eqnarray}
&& \Psi = \frac{\xbar \Psi}{r} + \frac{\Psi^{(1)}}{r^2} + o(r^{-2}), \\  \label{eq:BulkPsi}
&& 
\pi_\Psi =  \frac{1}{r^4} \pi^{(1)}_\Psi +
	o(r^{-4}). \label{eq:bcpiPsi}
\end{eqnarray}
(in Cartesian coordinates) and where $\lambda$ is a Lagrange multiplier for the constraint 
\be
\pi_\Psi \approx 0.
\ee
Since $\pi_\Psi$ is constrained to vanish on-shell, its precise decay at infinity can in fact be strengthened without eliminating classical solutions.

We have written explicitly the zero mode $\xbar \Theta_0$ of $\xbar \Theta$ among the arguments of the action  functional to emphasize that $S_H$ depends not just on $\xbar A_A = \d_A \xbar \Theta$, but on the full $\xbar \Theta$.  Shifts of $\Theta$ by constants will be seen to be symmetries.

The equations of motion that follow from the action are the original equations of motion for the original fields $A_i$, $\pi^i$, which imply
\be
\d_t (\xbar A_r + \xbar \Theta) = 0,
\ee
an equation which also follows by varying $\Psi$.  One also gets
\be
\d_t \xbar \Psi = 0
\ee
by varying with respect to the vector potential.  This equation is compatible with the equation obtained by varying with respect to $\pi_\Psi$, provided $\lambda \sim \frac{1}{r^2}$, which we shall assume.  One can in fact allow for a $\frac{1}{r}$-term in $\lambda$ if one introduces at the same time a non-vanishing surface Hamiltonian that reflects that the motion would then involve an improper gauge transformation for $\xbar \Psi$ (see Section \ref{sec:AsympSym2}), but we shall not do so here.  Finally, varying with respect to the Lagrange multipliers imply the constraints.

\subsection{Poincar\'e charges}
\label{sec:PoinCh1}

With the boundary modification of the symplectic charges, all Poincar\'e
transformations are canonical transformations with a well-defined generator.  
This generator can be written in terms of local diffeomorphisms
generators in the following way
\begin{gather}
	P_{\xi, \xi^i} = \int d^4x \left( \xi \mathcal H^{EM} + \xi^i
	\mathcal H^{EM}_i\right) + \mathcal B^{EM}_{(\xi, \xi^i)}, \label{eq:PoincEM000}\\
	\mathcal H^{EM} = -\Psi \d_i\pi^i -
	 A_i \nabla^i\pi_\Psi + 
	 \frac{1}{2\sqrt g} \pi_i \pi^i +
	 \frac{\sqrt g}{4} F_{ij} F^{ij},\\
	  \quad \mathcal H^{EM}_i = F_{ij} \pi^j - A_i \d_j \pi^j+ \pi_\Psi
	 \d_i \Psi,\\
	\mathcal B^{EM}_{\xi, \xi^i} = \oint d^3x \left(b \left(\xbar \Psi \xbar \pi^r
		 +\sqrt {\xbar\gamma} 
		\d_A \xbar \Phi \, \xbar D^A (\xbar A_r + \xbar \Theta) \right) +
        Y^A(\d_A \xbar \Phi \, \xbar \pi^r +
		\sqrt{\xbar\gamma} \,\xbar \Psi
	\d_A(\xbar A_r + \xbar \Theta))\right).
\end{gather}
As is well known, transformations of the fields under a symmetry are defined up to a gauge transformation in any gauge theory.  The  choice implicitly made in (\ref{eq:PoincEM000}) leads to a simple algebra.  For the kinematical transformations (spatial translations and rotations,) it is such that the action of these spatial symmetries on the fields is the ordinary Lie derivative, i.e.,
\be
\delta_{\xi^k} A_i = 
  \cL_{\xi^k} A_i, \quad  \delta_{\xi^k}\Psi =  \cL_{\xi^k} \Psi 
  \ee
 where $\Psi$ is a spatial scalar so that $\cL_\xi \Psi = \xi^k \d_k \Psi$ and where the spatial vector $\xi^k$ is given by (\ref{eq:FormOfW}). Note  that the Poincar\'e charges are invariant under shifts of $\xbar \Theta$ by constants.
 
To illustrate the derivation, consider the boosts ($\xi = b r$, $\xi^i = 0$).  The variations of all the canonical fields are given by
 \begin{eqnarray}
&& \delta_b A_i = \frac{\xi \pi_i}{\sqrt{g}} + \partial_i(\xi \Psi), \; \; \; \delta_b \pi^i =  \partial_m \left( F^{mi} \sqrt{g} \xi \right)+ \xi \nabla^i \pi_\Psi \\
&& \delta_b \Psi = \nabla^i \left(\xi A_i\right), \; \; \; \delta_b \pi_\Psi = \xi \partial_i \pi^i
 \end{eqnarray}
 from which one gets
\be
 \delta_b (\xbar A_r + \xbar \Theta) = \frac{b\xbar \pi^r}{\sqrt{\xbar \gamma}},  \; \; \; 
 \delta_b \xbar \Psi = \xbar D^A (b \partial_A \xbar \Phi)
\ee
upon appropriate choice of the constant that characterizes the ambiguity in $\xbar \Theta$.    Under these transformation, the symplectic two form 
 \be 
 \Omega = \Omega^{\rm{bulk}} + \Omega^{\rm{boundary}} 
 \ee
 with
 \be
 \Omega^{\rm{bulk}} =  \int d^4x \, \left( d_V \pi^i \, d_V A_ i + d_V \pi_\Psi \, d_V \Psi \right)
 \ee 
 and
 \be
 \Omega^{\rm{boundary}} = - \oint d^3x \, \sqrt{\xbar \gamma} \, d_V \left(\xbar A_r +  \xbar \Theta \right)\, d_V \xbar \Psi 
 \ee
 is invariant,  $\delta_b \Omega = d_V i_b \Omega = 0$, since one finds
 \be
 i_b \Omega^{\rm{bulk}} = - d_V \left( \int d^4x \xi \mathcal H^{EM} \right) - \oint d^3x \, b \, (d_V \xbar \pi^r) \, \xbar \Psi  - \oint d^3 x \, \sqrt{\xbar \gamma} \, b \xbar D^A (\xbar A_r + \xbar \Theta) \, d_V \partial_A \xbar \Phi
 \ee
 and
 \be
 i_b \Omega^{\rm{boundary}} = - \oint d^3 x \, b \, \xbar \pi^r \, d_V \xbar \Psi - \oint d^3 x \, \sqrt{\xbar \gamma}\, b \xbar D^A \left( d_V(\xbar A_r + \xbar \Theta)\right) \,  \partial_A \xbar \Phi
 \ee 
 so that
 \be
 i_b \Omega =  - d_V \left( \int d^4x \xi \mathcal H^{EM} + \oint d^3 x \, b \, \xbar \pi^r \, \xbar \Psi + \oint d^3 x \, \sqrt{\xbar \gamma}\, b \xbar D^A (\xbar A_r + \xbar \Theta) \, \partial_A \xbar \Phi\right)
 \ee
 The quantity of which $i_b \Omega$ is the exterior derivative is minus the canonical generator of the transformation, in agreement with (\ref{eq:PoincEM000}).
 
 Similarly, one finds for spatial rotations ($\xi = 0$, $\xi^r = 0$, $\xi^A =  Y^A $),
  \begin{eqnarray}
&& \delta_R A_i = \xi^j \d_j A_i  + \partial_i\xi^j A_j, \; \; \; \delta_R \pi^i =  \partial_m \left( \xi^m \pi^i \right)- \partial_m \xi^i \pi^m\\
&& \delta_R \Psi = \xi^i \partial_i \Psi, \; \; \; \delta_R \pi_\Psi =  \partial_i (\xi^i \pi_\Psi)
 \end{eqnarray}
 from which one gets
\be
 \delta_R (\xbar A_r + \xbar \Theta) = Y^A \partial_A (\xbar A_r + \xbar \Theta),  \; \; \; 
 \delta_R \xbar \Psi = Y^A \partial_A \xbar \Psi 
\ee
(with again a definite choice of the constant in $ \delta_R \xbar \Theta$). It follows that 
 \be
 i_R \Omega^{\rm{bulk}} = - d_V \left( \int d^4x \xi^i \mathcal H^{EM}_i  + \oint d^3x \, \xi^A \,  \xbar \pi^r \, \d_A \xbar \Phi \right)
 \ee
 and that 
 \be
 i_R \Omega^{\rm{boundary}} = - d_V \left(\oint d^3 x  \, \sqrt{\xbar \gamma}\, Y^A \d_A(\xbar A_r + \xbar \Theta) \, \xbar \Psi \right) 
  \ee 
yielding the above generator for spatial rotations.  Note that here, the bulk and boundary contributions to the symplectic form are separately invariant.

The computation of the time and spatial translations is simpler and leads to generators that have only a bulk piece.  This is because the relevant leading orders are invariant (it should be observed that for spatial translations $\delta_W \xbar A_r = - \partial_A \xbar \Phi \, \xbar D^A W$ and one can take $\delta_W \xbar \Theta = + \partial_A \xbar \Phi \, \xbar D^A W$ so that $\delta_W (\xbar A_r + \xbar \Theta) = 0$).

\subsection{Poincar\'e algebra}

In addition to the Poincar\'e symmetries, the theory is invariant under 
\be
\xbar \Theta \rightarrow \xbar \Theta + c
\ee
(everything else fixed), where $c$ is a constant. This transformation is generated by
\be
J = \oint d^3x \, \sqrt{\xbar \gamma} \, \xbar \Psi
\ee
since one has
\be
i_J \Omega = i_J \Omega^{\rm{boundary}} = -  \oint d^3x \, \sqrt{\xbar \gamma} \, c \, d_V \xbar \Psi = - d_V \left(cJ \right).
\ee
This quantity vanishes in $4$ spacetime dimensions where $\xbar \Psi$ is odd due to the parity conditions (and no ``naked'' $\xbar \Theta$ appears anyway).

One can easily compute the algebra of the various generators.  One finds:
\begin{gather}
	\label{eq:EMalgebraI}
	\{P_{\xi_1, \xi^i_1},P_{\xi_2, \xi^i_2}\} = P_{\hat\xi, \hat
	\xi^i}, \\
	\hat \xi = \xi_1^i \d_i \xi_2 - \xi_2^i \d_i \xi_1,
	\quad \hat \xi^i = \xi_1^j \d_j \xi_2^i - \xi_2^j \d_j \xi_1^i +
	g^{ij}(\xi_1 \d_j \xi_2 - \xi_1 \d_j\xi_2),
	\label{eq:EMalgebraII}
\end{gather}
which is the Poincar\'e algebra.

Furthermore, 
\be
\{P_{\xi, \xi^i},J \} = 0.
\ee

\section{Asymptotic symmetries}
\label{sec:AsympSym2}
\setcounter{equation}{0}

\subsection{Two sets of angle-dependent $u(1)$ symmetries}
We have already identified above an infinite set of asymptotic symmetries, which are gauge transformations parametrized by a gauge parameter that tends to a function $\xbar \epsilon$ on the $3$-sphere at infinity.  Contrary to the situation encountered in 3+1 dimensions all of these transformations are improper gauge transformations, i.e., non trivial symmetry transformations that generically change the physical state of the system, independently of their parity properties.  This brings a full angle dependent $u(1)$ set of asymptotic symmetries, with charge-generator equal (on-shell) to
\be
G[\epsilon] = \oint d^3 x \, \xbar \epsilon \, \xbar \pi^r.
\ee
 
The introduction of the surface field $\xbar \Psi$ at infinity brings in an independent second set of angle dependent $u(1)$asymptotic symmetries.  These are transformations that shift $\xbar \Psi$ by an arbitrary (time-independent) function $\xbar \mu$ of the angles.    They can be extended in the bulk as
\be
\delta_\mu \Psi = \mu, \delta_\mu (\hbox{anything else} )= 0
\ee
with
\be
\mu = \frac{\xbar \mu}{r} + o(r^{-1})
\ee
These are easily checked to leave the action invariant.
[One can include subleading terms in $\mu$, parametrized by arbitrary functions of the angles and of time, but these are proper gauge transformations leaving the action invariant provided $\lambda$ is transformed as $\delta_\mu \lambda = \d_t \mu$.] 

These transformations are canonical transformations with canonical generator given by
$$
G[\mu] = \int d^4x \mu \pi_\Psi + B
$$
where the surface term is necessary when $\mu \sim \frac{1}{r}$, even though the corresponding bulk constraint-generator is algebraic!  This is just as in $3+1$ dimensions.

One finds explicitly
\be
 i_\mu \Omega^{\rm{bulk}} = - d_V \left( \int d^4x \mu \pi_\Psi  \right), \; \; \; i_\mu \Omega^{\rm{boundary}} = -d_V \left(- \oint d^3 x \sqrt{\xbar \gamma} (\xbar A_r + \xbar \Theta ) \xbar \mu \right)
 \ee 
yielding as generator
\be 
G[\mu] = \int d^4x \mu \pi_\Psi - \oint d^3 x \sqrt{\xbar \gamma} (\xbar A_r + \xbar \Theta ) \xbar \mu 
\ee
an expression that reduces on-shell to
\be 
G[\mu] =  -\oint d^3 x \sqrt{\xbar \gamma} (\xbar A_r + \xbar \Theta ) \xbar \mu .
\ee
The generator $G[\mu]$, which generically does not vanish,  is  invariant under proper and improper gauge transformations.  One way to see this is to recall that $\d_A (\xbar A_r + \xbar \Theta ) = \xbar F_{rA}$, so that the gauge invariant asymptotic field strength $\xbar F_{rA}$ determines $\xbar A_r + \xbar \Theta$ up to a constant, and we fix the variation of the constant  in $\xbar \Theta$ so that $\delta \xbar A_r = - \delta \xbar \Theta$.

We thus come to the remarkable conclusion that the theory is invariant under angle-dependent $u(1) \oplus u(1)$ asymptotic symmetries, with generators given on-shell by
\be
G_{\mu, \epsilon} = \oint d^3 x \, \left( \xbar \epsilon \, \xbar \pi^r - \sqrt{\xbar \gamma} (\xbar A_r + \xbar \Theta ) \xbar \mu \right)
\ee

\subsection{Algebra}
\label{sub:Algebra}

The algebra of the global symmetries with themselves and the Poincar\'e generators are easily worked out.  One finds
\begin{gather}
	\{G_{\mu,\epsilon}, P_{\xi,\xi^i}\} = G_{\hat \mu, \hat
	\epsilon},\quad
	\{G_{\mu_1,\epsilon_1}, G_{\mu_2,\epsilon_2}\} = 0,\\
    \hat \mu = -\nabla^i(\xi\d_i \epsilon)-\xi^i\d_i \mu,\quad
    \hat\epsilon = -\xi \mu-\xi^i\d_i \epsilon. \label{eq:EMalgebra007}
\end{gather}
It follows from these equations that the algebra of the symmetries is a
semi-direct sum of the Poincar\'e algebra and the abelian algebra parametrized by $\xbar\mu$
and $\xbar\epsilon$. The action of the Poincar\'e subalgebra characterising
this semi-direct sum can be read off from \eqref{eq:EMalgebra007}:
\begin{gather}
    \delta_{(Y,b,T,W)}\xbar \mu = Y^A\d_A \xbar \mu + \xbar D_A(b \xbar D^A\xbar
	\epsilon), \quad \delta_{(Y,b,T,W)}\xbar \epsilon = Y^A \d_A \xbar
	\epsilon + b \xbar \mu.  \label{eq:EMalgebra008}
\end{gather}

One also finds by direct computation
\be
\{G_{\mu,\epsilon}, J \}=- \oint d^3 x \sqrt{\gamma} \xbar \mu,
\ee
i.e, a central charge appears in this Poisson bracket relation.

It should be observed that the even-parity component (under the antipodal map) of $\xbar \epsilon$ and the odd-partity component of $\xbar \mu$ transform into each other under Poincar\'e transformations. These are the only non-trivial components present in $4$ spacetime dimensions.  Similarly, the odd-parity component of $\xbar \epsilon$ and the even-parity component of $\xbar \mu$ transform also into each other.  These define proper gauge transformations in $4$ spacetime dimensions, but improper ones in $d>4$ spacetime dimensions when no parity condition need be imposed on the asymptotic fields.

\subsection{Time evolution}
Up to now, the field  $A_t$ is completely arbitrary.  A definite choice of $A_t$ amounts to a choice of which gauge transformation accompanies the time evolution generated by the standard Hamiltonian $\frac12 \int (E^2 + B^2)$.  It is convenient to choose
\be
A_t = \Psi
\ee
and we shall adopt this condition in the sequel.   This has a number of consequences:
\begin{itemize}
\item The behaviour of $A_t$ at spatial infinity takes the form
\begin{equation}
	A_t =  \frac{1}{r} \xbar A_t(x^A)+
	o(r^{-1}).
\end{equation}
i.e., the $O(1)$-piece $A_t^{(0)}(x^A)$ is set to zero, which means that no improper gauge transformation is added to the motion generated by $\frac12 \int (E^2 + B^2)$.  [This $O(1)$-piece can easily be re-inserted if needed, e.g., in discussing black hole thermodynamics.]
\item With the identification $A_t = \Psi$, the action reduces to
\begin{multline}
\label{eq:NonExtendedAction}
	S_H[A_i, \pi^i, A_t, \pi^0;  \lambda] = \int dt \left\{ \int d^4x \, 
		\pi^i \d_t
		A_i + \pi^0 \d_t A_t - \oint d^3x
		\, \sqrt {\xbar\gamma}\,  \left(\xbar A_r +  \xbar \Theta \right) \, \d_t \xbar A_t
		\right.\\
		\left. 	- \int d^4x \left(\frac{1}{2\sqrt g} \pi^i \pi_i + \frac{\sqrt
		g}{4} F^{ij} F_{ij} \right)  - \int d^4x \left( \lambda \pi^0 +  A_t \mathcal{G}  \right)\right\}.
\end{multline}
where $\pi^0 \equiv \pi_\Psi$.
\item The asymptotic equation $\d_t \xbar A_t =0$, which follows from $\d_t \xbar \Psi_t=0$, is asymptotically equivalent to the Lorenz gauge  $\partial^t A_t + \d^i A_i = 0$, since $\partial^t A_t = O(\frac1r)$, while $\d^i A_i = O(\frac{1}{r^2})$.
\end{itemize}
The action (\ref{eq:NonExtendedAction}) is in fact the action one obtains by direct application of the Dirac constrained Hamiltonian formalism. The primary constraints (here $\pi^0= 0$) are enforced explicitly with their own Lagrange multipliers (here $\lambda$).   The secondary constraints (here $\mathcal{G} =0$) follow from the preservation in time of the primary constraints.  The parameters of the gauge transformations generated by $\pi^0$ and  $\mathcal{G}$ are related in this ``unextended formulation'' by $\dot{\epsilon} = \mu$ (more in \cite{Henneaux:1990au})\footnote{The Lorenz gauge involves the time derivative of $A_t$ and thus is invariant under gauge transformations with a gauge parameter $\epsilon$ that obeys a second order equation with respect to time.  This means that on an initial slice, the gauge parameters $\mu= \dot{\epsilon}$ and $\epsilon$ are completely free: they obey no gauge condition and there is no gauge fixing to be preserved.}.  

For that reason, the introduction of $\Psi$ is not really the introduction of a new physical degree of freedom. It corresponds rather to the explicit recognition that the $O(\frac1r)$-part of $A_0$ is not pure gauge, which must therefore be kept. It would be incorrect to set it equal to zero since this would require the use of an improper gauge transformation.  That shifts of $\xbar A_t$ are not proper gauge transformations reflects itself through the fact that the corresponding generator is $\int d^4x \mu \pi^0- \oint d^3 x \sqrt{\xbar \gamma} (\xbar A_r + \xbar \Theta ) \xbar \mu $ and does not generically vanish, even on-shell. The advantage of including the $\Psi$ field is that this important property is clearly put to the foreground.  But once understood, one can equivalently work with the more familiar action (\ref{eq:NonExtendedAction}).  This action reproduces the standard Maxwell action (up to a surface term) with $A_\mu$ as sole dynamical variables if one performs the Legendre transformation in reverse, back to the Lagrange formalism.

\section{More than $5$ spacetime dimensions}
\label{sec:More5}
\setcounter{equation}{0}

The extension of the formalism to other  dimensions higher than $4$ is direct and remarkably ``uneventful''. There is in particular no difference between odd and even spacetime dimensions.  We give below the relevant formulas.

\subsection{Action, boundary conditions, Poincar\'e generators}

The action generalizing (\ref{eq:NonExtendedAction}) is
\begin{multline}
\label{eq:NonExtendedActionHigherD}
	S_H[A_i, \pi^i, A_t, \pi^0, \xbar \Theta_0;  \lambda] = \int dt \left\{ \int d^{d-1}x \, 
		\pi^i \d_t
		A_i + \pi^0 \d_t A_t - \oint d^{d-2}x
		\, \sqrt {\xbar\gamma}\,  \left(\xbar A_r +  (d-4) \xbar \Theta \right) \, \d_t \xbar A_t
		\right.\\
		\left. 	- \int d^{d-1}x \left(\frac{1}{2\sqrt g} \pi^i \pi_i + \frac{\sqrt
		g}{4} F^{ij} F_{ij} \right)  - \int d^{d-1}x \left( \lambda \pi^0 +  A_t \mathcal{G}  \right)\right\}.
\end{multline}

The asymptotic conditions are (at a given time)
\begin{gather}
	A_r = \partial_r \Phi +\frac{1}{r^{d-3}} \xbar A_r + \frac{1}{r^{d-2}} A^{(1)}_r +
	o(r^{{2-d}}),\quad
	\pi^r = \xbar \pi^r + \frac{1}{r} \pi^{(1)r} +
	o(r^{-1}),\\
	A_A = \partial_A  \Phi + \frac{1}{r^{d-4}}\d_A\xbar \Theta + \frac{1}{r^{d-3}} A^{(1)}_A +
	o(r^{3-d}),\quad
	\pi^A =  \frac{1}{r^2} \pi^{(1)A} +
	o(r^{-2}),\\
	A_t =  \frac{1}{r} \xbar A_t(x^A)+ \frac{1}{r^2} A_t^{(2)} + \cdots  \frac{1}{r^{d-4}} A_t^{(d-4)} +
	o(r^{4-d})  \\
	\pi^0 = \frac{1}{r} \pi_{(1)}^0 + o(r^{-1}) \\
	\Phi = \xbar \Phi + \frac{1}{r} \Phi^{(1)} + \cdots + \frac{1}{r^{d-5}} \Phi^{(d-5)}
\end{gather}
and involves no parity conditions ($d >4$).  The Lagrange multiplier $\lambda$ fulfills
\be
\lambda = O(\frac{1}{r^2}).
\ee
We stress that we have given the asymptotic behaviour of the fields, in particular of the density $\pi^0$,  in polar coordinates (in cartesian coordinates, $\pi^0 \sim r^{1-d}$).  

The equations of motion imply
\be
\d_t \xbar A_t = 0
\ee
which is equivalent to
\be
\partial^\mu A_\mu = O(\frac{1}{r^2})
\ee
(instead of the expected $O({\frac1r})$ that would follow from a generic time-dependence).  The Lorenz gauge is  therefore fulfilled asymptotically.

The Poincar\'e generators are
\begin{gather}
	P_{\xi, \xi^i} = \int d^{d-1}x \left( \xi \mathcal H^{EM} + \xi^i
	\mathcal H^{EM}_i\right) + \mathcal B^{EM}_{(\xi, \xi^i)}, \label{eq:PoincEMHigh}\\
	\mathcal H^{EM} = -A_t\d_i\pi^i -
	 A_i \nabla^i\pi^0 + 
	 \frac{1}{2\sqrt g} \pi_i \pi^i +
	 \frac{\sqrt g}{4} F_{ij} F^{ij},\\
	  \quad \mathcal H^{EM}_i = F_{ij} \pi^j - A_i \d_j \pi^j+ \pi^0
	 \d_i A_t,\\
	\mathcal B^{EM}_{\xi, \xi^i} = \oint d^{d-2}x \left(b \left(\xbar A_t \xbar \pi^r
		 +\sqrt {\xbar\gamma} 
		\d_A \xbar \Phi \, \xbar D^A (A_r + (d-4) \Theta) \right) \right) \nonumber \\
        \hspace{2cm} + \oint d^{d-2}x \left( Y^A(\d_A\xbar
        \Phi \, \xbar \pi^r +
		\sqrt{\xbar\gamma} \,\xbar A_t
    \d_A(\xbar A_r + (d-4) \xbar \Theta))\right).
\end{gather}

There is also the symmetry $\xbar \Theta \rightarrow \xbar \Theta + c$ generated by 
\be
J = \oint d^{d-2} x \sqrt{\xbar \gamma} \, \xbar A_t
\ee

\subsection{Asymptotic symmetries}

The theory has gauge symmetries generated by
\begin{eqnarray}
\label{eq:NoetherChargeGenerator1}
G_{\epsilon, \mu} &=& \int d^{d-1}x \, \left( \epsilon \, \mathcal{G} + \mu \, \pi^0 \right) + \oint d^{d-2}x \, \left(\xbar \epsilon \, \xbar \pi^r -  \sqrt{\xbar \gamma} \,  \xbar \mu \, (\xbar A_r + (d-4) \xbar \Theta )  \right) \\ &\approx&
	\oint d^{d-2}x \, \left(\xbar \epsilon \, \xbar \pi^r -  \sqrt{\xbar \gamma}\,  \xbar \mu \,(\xbar A_r + (d-4) \xbar \Theta )  \right)
	\label{eq:NoetherChargeGenerator2}
\end{eqnarray}
with
\begin{equation}
	\epsilon = \xbar \epsilon(x^A)  + \frac1r \epsilon^{(1)} + O(\frac{1}{r^2})
	 \label{eq:GaugeParHighD1}
\end{equation}
and
\be
\mu = \frac{\xbar \mu (x^A)}{r} +  O(\frac{1}{r^2}) \label{eq:GaugeParHighD2}
\ee
The action is invariant provided the following (non-independent) conditions hold: (i) $\d_t \xbar \epsilon = 0$; (ii) $\d_t \xbar \mu = 0$; (iii) $\d_t \epsilon = \mu$.  

The unwritten $O(\frac{1}{r^2})$-terms in (\ref{eq:GaugeParHighD1}) are arbitrary functions of time and define proper gauge transformations.   The corresponding $O(\frac{1}{r^2})$-terms in (\ref{eq:GaugeParHighD2}) are determined by the condition $\mu = \dot{\epsilon}$ (in the unextended canonical formalism considered here).  One could use these proper gauge transformations to eliminate all terms in $\Phi$ but the leading one, but this is not necessary. 

By contrast, the leading terms parametrized by $\xbar \epsilon(x^A)$ and $\xbar \mu (x^A)$, which are arbitrary time-independent functions of the angles restricted by no parity condition ($d >4$) define improper gauge transformations.  The asymptotic symmetries are thus  the direct sum of two independent sets of angle-dependent $u(1)$ transformations.  Note that $ \epsilon^{(1)}$ is determined by the condition $\d_t \epsilon = \mu$ to be $ \epsilon^{(1)} = t \xbar \mu (x^A)$ and that the generator (\ref{eq:NoetherChargeGenerator1})-(\ref{eq:NoetherChargeGenerator2}) is just the corresponding Noether charge.

The algebra of the generators of asymptotic symmetries and Poincar\'e transformations is the same as in Subsection \ref{sub:Algebra}.

We close this section by observing that $\partial^2_t \epsilon = O(\frac{1}{r^2})$ is equivalent to $\Box \epsilon = O(\frac{1}{r^2})$.  Now, the condition $\Box \epsilon = 0$ defines the residual gauge freedom in the Lorenz gauge.  The asymptotic symmetries are therefore residual gauge transformations of the asymptotic Lorenz gauge.  What makes them non-trivial symmetry transformations, however, is not that they are residual gauge transformations for some gauge conditions, but that their canonical generator has a non-vanishing value\footnote{One can impose the Lorentz gauge to all orders.  In that case, the residual gauge transformations fulfill $\Box \epsilon = 0$ to all orders.  The general solution to that equation that starts like $\epsilon = \xbar \epsilon + \frac{t \xbar \mu}{r} + \sum_{n \geq 2} \frac{\epsilon^{(n)}}{r^n}$ will have generically $\epsilon^{(n)} \sim t^n$.  Each $\epsilon^{(n)}$ will involve two new integration ``constants''  $C_N(x^A)$ (functions of the angles), but these integration ``constants''  correspond to pure gauge transformations, even though associated with residual gauge transformations. }.

\section{Connection with null infinity}
\label{sec:NullInf}

The formulation on Cauchy hyperplanes in Mikowski spacetime ${\mathbb R}^{d-1,1} \simeq {\mathbb R} \times {\mathbb R}^{d-1}$ is complete and fully specifies the system.   It is self-contained and sufficient to answer all dynamical questions, including asymptotic ones for $t \rightarrow \pm \infty$ or any other limit.   

\subsection{Hyperbolic coordinates}

It is of interest in particular to derive from the present formulation the behaviour of the fields as one goes to null infinity.  To that end, we first integrate the equations in hyperbolic coordinates \cite{Ashtekar:1978zz}, were the metric is
given by
\begin{equation}
	ds^2 = d\eta^2 + \eta^2 h_{ab} dx^a dx^b, \quad h_{ab} dx^a dx^b =
	-\frac{1}{(1-s^2)^2} ds^2 + \frac{\xbar \gamma_{AB}}{1-s^2} dx^A dx^B.
\end{equation}
Here,  $h_{ab}$ is the metric on the $(d-1)$ hyperboloid.
The explicit change of variables is
\be 
\eta = \sqrt{-t^2 + r^2}, \; \; \; s = \frac{t}{r}
\ee
The hyperbolic patch covers the region $r > \vert t \vert$.
The inverse transformation reads
\begin{equation}
    t = \eta \frac {s}{\sqrt{1-s^2}}, \quad r = \eta \frac
    {1}{\sqrt{1-s^2}}. 
    \end{equation}
The use of hyperbolic coordinates in this context is known to be very useful    \cite{Ashtekar:1978zz,BeigSchmidt,Beig:1983sw,Compere:2011ve,Fried1,Friedrich:1999wk,Friedrich:1999ax}.

The hypersurface $s=0$ in hyperbolic coordinates, on which $\eta = r$, coincides with the hypersurface $t=0$ and is therefore a Cauchy hypersurface on which we have already studied the behaviour of the fields at infinity.  This leads to  the following fall-off for the electromagnetic potential
\begin{equation}
	A_\eta = \partial_\eta \Phi + \xbar A_\eta \eta^{-d+3} + O(\eta^{-d+2}), \qquad A_a = \d_a 
	\Phi + \xbar A_a \eta^{-d+4} + O(\eta^{-d+3}),
\end{equation}
with field strengths taking the asymptotic form
\begin{equation}
	E_b = F_{b\eta} = \eta^{-d+3} \Big(\d_b \xbar A_\eta + (d-4) \xbar
	A_b\Big) + O(\eta^{-d+2}), \quad F_{ab} = \eta^{-d+4} \xbar F_{ab} +
		O(\eta^{-d+3}).
\end{equation}
 Here,
\be
\Phi = \xbar \Phi + \cdots + \frac{1}{\eta^{d-5}} \Phi^{(d-5)}.
\ee
Note that the temporal component $A_s$ of the vector potential in hyperbolic coordinates is boosted by one power of $\eta$ with respect to $A_t$.  As above, we will also assume $\xbar F_{ab} = 0$ (i.e., $F_{sA} = 0$ and $F_{AB} = 0$), which means that there exists a boundary
field $\Theta$ such that $\xbar A_a = \d_a \Theta$. 

The equations of motion are given by
\begin{equation}
    \mathcal D^a E_a = 0, \qquad \eta^{d-5} \mathcal D^b F_{ba} -
    \d_\eta(\eta^{d-3} E_a) = 0
\end{equation}
with $E_b \equiv F_{b\eta}$.  Here, $\mathcal D_a$ is the covariant derivative associated with the hyperboloid metrix $h_{ab}$, while latin indices are raised or lowered with $h^{ab}$ and $h_{ab}$, respectively.

Although the asymptotic analysis only needs the leading orders, it is instructive to assume that the asymptotic behaviour of the fields can be expressed in terms of an expansion in $\eta^{-1}$:
\begin{equation}
    \label{eq:EOMexp}
    E_b = F_{b\eta} = \sum_{k\geq 0} \eta^{-d + 3 - k} E_b^{(k)}, \qquad
    F_{ab} = \sum_{k \geq 0} \eta^{-d + 4 - k} F_{ab}^{(k)}, \qquad F_{ab}^{(0)} = 0.
\end{equation}
This is because the equations of motion imply a decoupling order by order.  Indeed, one gets
\begin{equation}
    \mathcal D^a E^{(k)}_a = 0, \qquad 
    \mathcal D^b F^{(k)}_{ba} + k E^{(k)}_a = 0,
\end{equation}
Similarly, the  Bianchi identity becomes
\begin{equation}
    (d-4+k) F_{ab}^{(k)} = \d_a E^{(k)}_b - \d_b E^{(k)}_a, \qquad
    \d_{[a} F^{(k)}_{bc]} = 0.
\end{equation}
This leads to the second order equation
\begin{equation}
    \label{eq:maintexteomE}
    \mathcal D^b \mathcal D_b E^{(k)}_a - (d-2) E^{(k)}_a +
    k(d-4+k) E^{(k)}_a= 0.
\end{equation}
containing $E^{(k)}_a$ only.

We now focus on the component $E_s^{(k)}$ of the
curvature. As in the 4d case, this is the most interesting one as it
carries the information about the charges. The analysis of the other components of the field strength can be
found in appendix \ref{app:solEOM}.

Using the fact that
$E_a^{(k)}$ is divergence-less, the component $a=s$ of equation
\eqref{eq:maintexteomE} reads
\begin{equation}
    (1-s^2) \d_s^2 E^{(k)}_s + (d-6)s \d_s  E^{(k)}_s + (d-4) E^{(k)}_s -
    D_A D^A E^{(k)}_s  - (1-s^2)^{-1}k (d-4+k) E^{(k)}_s=0.
\end{equation}
Making then the following rescaling of the components:
\begin{equation}
    \Xi^{(k)} = (1-s^2)^{\frac k 2 } E_s^{(k)},
\end{equation}
leads to
\begin{equation}
    (1-s^2) \d_s^2 \Xi^{(k)} + (d-6+2k)s \d_s  \Xi^{(k)} -
    (k-1)(k+d-4) \Xi^{(k)} - D_A D^A \Xi^{(k)} =0.
\end{equation}
Using a basis of spherical harmonics for the $(d-2)$-sphere (see Appendix \ref{sub:SpheHar}), we get
\begin{gather}
    \Xi^{(k)} = \sum_{l,m} \Xi_{lm}^{(k)}(s) Y_{lm}(x^A), \qquad D^AD_A
    Y_{lm} = -l(l+d-3) Y_{lm}, \\
    (1-s^2) \d_s^2 \Xi_{lm}^{(k)} + (d-6+2k)s \d_s  \Xi_{lm}^{(k)} -
    (k-1)(k+d-4) \Xi_{lm}^{(k)} + l(l+d-3) \Xi_{lm}^{(k)} =0.
\end{gather}
With $\lambda = k + \frac{d-3} 2$ ($=$ integer or half-integer) and $n = l-k$, we can rewrite this
equation as
\begin{equation}
    (1-s^2) \d_s^2 \Xi_{lm}^{(k)} + (2\lambda-3)s \d_s  \Xi_{lm}^{(k)} +
    (n+1)(2\lambda + n -1) \Xi_{lm}^{(k)} =0. \label{eq:UltraKey}
\end{equation}
This equation is exactly the same as the equation that arises in the discussion of the asymptotics of the scalar field discussed in \cite{Henneaux:2018mgn}, where the necessary mathematical background was recalled.  For the sake of completeness of this paper, the needed results are reproduced in Appendix \ref{sub:Ultra}.

As shown in Appendix \ref{sub:Ultra}, the general solution
to the equation (\ref{eq:UltraKey}) takes the form:
\begin{equation}
    \Xi_{lm}^{(k)} = \Xi_{lm}^{P(k)}
    \tilde P^{(\lambda)}_n (s) + \Xi_{lm}^{Q(k)}
    \tilde Q^{(\lambda)}_n (s).
\end{equation}
In the limit $s\to \pm 1$, the resulting function $\Xi^{(k)}_{lm}$ tends
to a constant. When $l\ge k$, the contribution from the $P$-branch is
sub-leading.

The complete analysis given in appendix \ref{app:solEOM} leads to similar expressions
for the other components of the curvature. An interesting feature that emerges from the analysis is that the divergence-less
condition on $E_a^{(k)}$ implies that most of the zero-modes of
$E_s^{(k)}$ are zero:
\begin{equation}
    \Xi_{00}^{P(k)} = 0, \quad 
    \Xi_{00}^{Q(0)} = 0, \quad 
    \Xi_{00}^{Q(k)} = 0, \quad \forall k>0.
\end{equation}

\subsection{Behaviour at null infinity}

 In order to make
the link with null infinity, it is necessary to rescale the radial
coordinate $\rho = \eta \sqrt{1-s^2}$ \cite{Fried1,Friedrich:1999wk,Friedrich:1999ax}. Indeed, the hyperbolic coordinates badly describe the limit to null infinity: the pair $(s, \eta)$ always tends to $(1, \infty)$ on all outgoing null geodesics $t = r + b$, no matter what $b$ is.  By contrast, the coordinate $\rho$ tends to $2 \vert b \vert$ and can distinguish between the various null  geodesics.  The detailed procedure on how to derive the evolution of the  Cauchy data as one goes  to null infinity is discussed at length in \cite{Fried1,Friedrich:1999wk,Friedrich:1999ax}, to which we refer the reader. Note the use of the radial coordinate $\frac{1}{\rho}$ there, which goes to zero in the null infinity limit. 

A relevant quantity is then the
electric density:
\begin{flalign}
    \sqrt{-g} F^{\rho s} & = \sqrt {\xbar \gamma} \rho^{d-3}(1-s^2)^{-d + \frac 7 2}
    E_{s}(\rho, s, x^A) \nonumber \\& 
    = \sqrt {\xbar \gamma} (1-s^2)^{-\frac d 2 + 2} \sum_k \rho^{- k}\Xi^{(k)}.
\end{flalign}
Let us look at the leading term $k=0$, separating the two branches, we get
\begin{equation}
    \sqrt{-g} F^{\rho s} 
    = \sqrt {\xbar \gamma} \sum_{lm} \Xi_{lm}^{P(0)}
     (1-s^2)^{-\frac d 2 + 2}\tilde P^{(\frac{d-3}{2})}_l Y_{lm}+ 
    \sqrt {\xbar \gamma} (1-s^2)^{-\frac d 2 + 2} \sum_{lm}\Xi_{lm}^{Q(0)}
     \tilde Q^{(\frac{d-3}{2})}_l Y_{lm} + O(\rho^{-1}).
\end{equation}
In the limit $s\to \pm 1$, the first term will go to a constant while the
second one will diverge. Surprisingly, when we look at the sub-leading
contributions in $\rho^{-1}$, all contributions from $Q^{(\lambda)}_n$
along with the contributions from $P^{(\lambda)}_n$ with
$n=l-k<0$ will diverge in the same way when $s\to\pm 1$. The rest of the
contribution from the $P$'s will be sub-leading when $s\to\pm 1$:
\begin{equation}
    \sqrt{-g} F^{\rho s} 
    = \sqrt \gamma (1-s^2)^{-\frac d 2 + 2} \sum_k\sum_{lm} \rho^{- k} \Big (\Xi_{lm}^{P(k)}
    \tilde P^{(k+\frac{d-3}2)}_{l-k} + \Xi_{lm}^{Q(k)}
    \tilde Q^{(k+\frac{d-3}2)}_{l-k} \Big)Y_{lm}.
\end{equation}

Performing the change of variables leading to standard retarded null coordinates,
\begin{equation}
    u = \eta \frac {s-1}{\sqrt{1-s^2}}, \qquad r = \eta
    \frac{1}{\sqrt{1-s^2}},
\end{equation}
so that 
\begin{equation}
    s = 1 + \frac u r, \qquad \rho = -2u - \frac {u^2}{r},
\end{equation}
($u<0$), we get
\begin{equation}
    F_{ur} = \frac{1-s^2}{\eta} E_s = \rho^{-1} (1-s^2)^{\frac{3}{2}}
    E_s(\rho,s,x^A).
\end{equation}
Using the expansion we obtained above, this leads to
\begin{flalign}
    F_{ur}
    &= \sum_k\rho^{2-d- k} (1-s^2)^{\frac d 2} \sum_{lm} \Xi_{lm}^{P(k)}
    \tilde P^{(\lambda)}_n (s)Y_{lm} 
    \nonumber \\ & \qquad 
    +\sum_k \rho^{2 - d- k} (1-s^2)^{\frac d 2} \sum_{lm}\Xi_{lm}^{Q(k)}
    \tilde Q^{(\lambda)}_n (s)Y_{lm}.
\end{flalign}
Substitution of $s$ and $\rho$ then yields
\begin{flalign}
    F_{ur}
    &= r^{2-d} \sum_{lm} \Xi_{lm}^{P(0)}
    \lim_{s\to 1}P^{(\frac{d-3}{2})}_n Y_{lm} + O(r^{1-d})    \nonumber \\ & \qquad 
    + r^{-\frac
    d 2} (-2u)^{2 - \frac d 2}\sum_k (-2u )^{- k}\lim_{s\to
    1}\Big(\sum_{l\ge 0,m}
    \tilde Q^{(\lambda)}_n \Xi_{lm}^{Q(k)}Y_{lm} + \sum_{k> l\ge 0,m}
    \tilde P^{(\lambda)}_n \Xi_{lm}^{P(k)}Y_{lm}\Big)+ O(r^{-\frac d 2-1}).
\end{flalign}
We can thus conclude that in the $r^{-1}$ expansion,
the various orders in $\eta^{-1}$ of the $Q$ branch combined with some
spherical harmonic component of the $P$ branch will build a 
function of $u$ at order $r^{-\frac d 2}$ while the rest of the $P$ branch will only
contribute to a $u$-constant term at order $r^{2-d}$. It should be noted that whereas there is no fractional powers of $\eta$ in hyperbolic coordinates, fractional powers of $r$ appear in null coordinates, in the case of odd spacetime dimensions. 

The leading term in $r^{2-d}$ represents the Coulomb part, while the leading term in $r^{-\frac{d}{2}}$ represents the radiation part.   Our expansion in null coordinates is in agreement with the results of \cite{Satishchandran:2019pyc}.

In $4$ spacetime dimensions,  the $k=0$ contribution of the
$Q$ branch acquires a leading logarithmic divergence \cite{Henneaux:2018gfi,Henneaux:2018hdj}. Assuming that this
contribution is absent, i.e. $\Xi^{Q(0)}_{lm} = 0$, then the above
expansion is still valid for $d=4$. In that case, the remaining $Q$
contributions and the leading $P$ contribution appear at the same order
$r^{-2}$.  There is no such singularity in $d>4$ dimensions and so no need to remove the $k=0$ contribution of the $Q$ branch. 

Although this is not necessary, one might be tempted to remove the $Q$ branches at all orders along with all the $P$ branches 
contribution with $k\ge l$, in order to get rid of the $r^{-\frac{d}{2}}$-terms. The only component in $F_{ur}$ is then
at $r^{2-d}$, which encodes as we mentioned the Coulomb part associated with the parity
even branch at the leading order at spatial infinity: the $P$-branch for
$k=0$.  However, when $d>4$, this contribution is constant along scri and it
will not build a generic function of $u$. From our analysis, we see that
the relevant order is $r^{-\frac d 2}$, which must be kept. A curiosity in $4$ spacetime dimensions is that
this is also the order at which the Coulomb contribution appears. 

\subsection{Generalized matching conditions}

To come back to the case of $5$ spacetime dimensions, our
analysis implies the following asymptotics at null infinity for $d=5$:
\begin{equation}
    F_{ur} = A(u,x^A) r^{-\frac 5 2 } + C(x^A) r^{-3} + \ldots, \quad
    A(u,x^A) = \frac 1 {\sqrt{-u}} \xbar A(x^A) +
   O\Big((-u)^{-\frac 3 2}\Big).
\end{equation}
 Both $C$ and $\xbar A$  have opposite matching conditions since $\xbar A$ comes
from the $Q^{(0)}$ branch while  $C$ comes from the $P^{(0)}$
branch.  

The generalized matching conditions will thus involve both even and odd matchings ($P$ is even under the sphere antipodal map combined with $s \rightarrow -s$ but $Q$ is odd). These matching conditions are associated with different powers of $r$.   The same procedure works in higher dimensions.  In $4$ dimensions where one must set $\xbar A =0$, there is only the even component \cite{Henneaux:2018gfi,Henneaux:2018hdj}.

\subsection{Gauge transformations}

In hyperbolic coordinates where the $s$-time is boosted by one power of $\eta$, the two types of improper gauge transformations appear at the same order $O(\eta^0)$.  Technically, this follows from the fact that $s = \frac{t}{r}$, so that terms such as $\left(\frac{t}{r} \right)^k$, which decrease as $r^{-k}$ on constant $t$-slices, behave as $s^k$, i.e., as constants on constant $s$-slices.

Gauge transformations are generated by parameters $\epsilon$ of the form
\begin{equation}
	\epsilon(\eta,x^a) = \xbar \epsilon(x^a) 
	+ O(\eta^{-1}), \qquad \mathcal D^a \mathcal D_a \xbar \epsilon = 0
\end{equation}
where, as we have seen, the Lorenz gauge holds at infinity.  The generator is given on-shell by
\begin{equation}
	\mathcal Q_\epsilon \approx \oint (d^{d-2}x) \sqrt{-h} h^{sb}
	\left[\xbar E_b \xbar \epsilon
	- \Big(\xbar A_\eta + (d-4) \Theta\Big) \d_b \xbar \epsilon\right]
\end{equation}
where the first term,associated with the gauge parameter undifferentiated with respect to time and linked with the flux of the electric field $E_s \sim F_{\eta s}$, generates the first type of improper gauge transformations, while the second one, associated with the $s$-derivative of the gauge parameter, generates the second type of improper gauge transformations.  The  generator $\mathcal Q_\epsilon$ is conserved.

The equation for  $\xbar \epsilon$ can be rewritten as
\begin{equation}
	(1-s^2) \d_s^2 \xbar \epsilon + (d-4) s \d_s \xbar \epsilon - D_A D^A
	\xbar \epsilon = 0,
\end{equation}
which leads to
\begin{equation}
	\xbar \epsilon = \sum_{l,m} \epsilon_{l,m}(s) Y_{l,m}(x^a), \qquad 
	(1-s^2) \d_s^2 \epsilon_{l,m} + (d-4) s \d_s \epsilon_{l,m} 
	+ l (l + d -3)\epsilon_{l,m} = 0.
\end{equation}
This corresponds to equation \eqref{eq:alternJacobieq} with 
$\lambda = \frac 1 2 (d-1)$ and $n = l-1$. For $l>0$, we can write the
solution as
\begin{equation}
	\epsilon_{l,m} = 
	(1-s^2)^{\frac d 2-1 }\left(\epsilon^P_{l,m} P^{(\frac d 2 - \frac 1
		2)}_{l-1}(s) + \epsilon^Q_{l,m}
	Q^{(\frac d 2 - \frac 1 2)}_{l-1}(s)\right).
\end{equation}
The term for $l = 0$ is given by
\begin{equation}
	\epsilon_{0,0}(s) = \epsilon^P_{0,0}\int_0^s (1-x^2)^{\frac d 2 -2}dx +\epsilon^Q_{0,0}. 
\end{equation}
As for the radial electric field in the previous section, we have two branches
of solutions for $\xbar \epsilon$ characterised by their behaviour under 
parity. The solution parametrised by $\epsilon^Q_{l,m}$ is odd while the one
parametrised by $\epsilon^P_{l,m}$ is even. The odd solution tends to a finite
function on the sphere in the limit $s\to \pm 1$ while the even solution tends
to zero except for the zero mode $l=0$.

At first sight, the behaviour of these $u(1)$
transformations in the limit $s\to 1$ might be
suprising. In particular, it would seem that the $P$ branch falls off too fast
and will lead to a zero charge close to null infinity.
Let us focus on $d=5$ and look at the behaviour of the charges in the limit
$s\to 1$. From our previous analysis, we have
\begin{equation}
    \xbar E_s = \d_s(\xbar A_\eta + \Theta), \qquad \mathcal D^a E_a =
    \mathcal D^a \mathcal D_a(\xbar A_\eta + \Theta) = 0.
\end{equation}
The combination $\xbar A_\eta + \Theta$ satisfies the same equation as
$\xbar\epsilon$ and takes the general form
\begin{equation}
    \xbar A_\eta + \Theta = \sum_{l, m}
	\left(\Theta^P_{l,m} \tilde P^{(2)}_{l-1}(s) + \Theta^Q_{l,m}
    \tilde Q^{(2)}_{l-1}(s)\right)Y_{l,m},
\end{equation}
where $\Theta^P_{lm}$ and $\Theta^Q_{lm}$ can be expressed in terms of
$\Xi^{(0)P}_{lm}$ and $\Xi_{lm}^{(0)Q}$.

In the limit $s\to 1$, the ultraspherical functions with $n\ge 0$ behave as:
\begin{flalign}
    \tilde P_{n}^{(2)} & =  (1-s^2)^{\frac 3 2} \left[P_n^{(2)0} + P_n^{(2)1}(1-s) +
    O\Big((1-s)^2\Big) \right],\\
    \tilde Q_{n}^{(2)} & = Q_n^{(2)0} + Q_n^{(2)1}(1-s) +
    O\Big((1-s)^2\Big).
\end{flalign}
Taking into account these asymptotic behaviours, we can expand the charge
in the neighbourhood of future null infinity:
\begin{flalign}
    \mathcal Q_\epsilon & = -\sum_{l>0,m}(1-s^2)^{- \frac 1 2}
    \left[\d_s\Big(\Theta^P_{l,m} \tilde P^{(2)}_{l-1}(s) + \Theta^Q_{l,m}
    \tilde Q^{(2)}_{l-1}(s)\Big) \Big(\epsilon^P_{l,m} \tilde P^{(2)}_{l-1}(s) +
    \epsilon^Q_{l,m}
    \tilde Q^{(2)}_{l-1}(s)\Big)
	- \Theta \leftrightarrow \epsilon\right]\nonumber\\
    & \qquad - \Theta^P_{0,0} \epsilon^Q_{0,0} + \Theta^Q_{0,0} \epsilon^P_{0,0}\\
    & = -\sum_{l>0,m}(1-s^2)^{- \frac 1 2}
    \Big[\d_s\Big(\Theta^Q_{l,m} Q^{(2)0}_{l-1}+\Theta^Q_{l,m}
    Q^{(2)1}_{l-1}(1-s) + \Theta^P_{l,m}
    P^{(2)0}_{l-1}(1-s^2)^{\frac 3 2} + O\Big((1-s)^2\Big)\Big)\nonumber \\
    &\qquad \qquad \Big(\epsilon^Q_{l,m} Q^{(2)0}_{l-1}+\epsilon^Q_{l,m}
    Q^{(2)1}_{l-1}(1-s) + \epsilon^P_{l,m}
    P^{(2)0}_{l-1}(1-s^2)^{\frac 3 2} + O\Big((1-s)^2\Big)\Big)
    - \Theta \leftrightarrow \epsilon\Big]\nonumber\\
    & \qquad - \Theta^P_{0,0} \epsilon^Q_{0,0} + \Theta^Q_{0,0}
    \epsilon^P_{0,0}\nonumber\\
    & = \sum_{l>0,m}
    \Big[3(\epsilon^Q_{l,m} \Theta^P_{l,m} - \epsilon^P_{l,m} \Theta^Q_{l,m}) Q^{(2)0}_{l-1}
    P^{(2)0}_{l-1} + O\Big((1-s)^{\frac 1 2}\Big)\Big)\Big]
    - \Theta^P_{0,0} \epsilon^Q_{0,0} + \Theta^Q_{0,0} \epsilon^P_{0,0}.
\end{flalign}
As expected, the various modes pair up and we do recover the two
$u(1)$'s.

\section{Conclusions}
\label{sec:Conclusions}

One striking result of our analysis is the emergence of an asymptotic symmetry algebra that is the direct sum of two angle-dependent $u(1)$'s for $d>4$, instead of just one angle-dependent $u(1)$.  

This somewhat unexpected result can be understood as follows.  It is customary to say that in electromagnetism (and for that matter, also in gravity), ``the gauge symmetry strikes twice''.  That is, both  temporal and longitudinal components of $A_\mu$ are removed by the single gauge invariance $\delta_\epsilon A_\mu = \d_\mu \epsilon$.   This property follows from the fact that the gauge symmetry involves both $\epsilon$ and $\dot{\epsilon}$, which are independent at any given time, so that invariance of the physical states (defined on a spacelike surface) under gauge transformations yields two conditions, one related to $\epsilon$  and one related to $\dot{\epsilon}$.   Classically, this is reflected in the presence of two first class constraint-generators per space point, $\pi^0 \approx 0$ and $\mathcal{G} \approx 0$ \cite{Henneaux:1990au}.

The ``obvious'' angle-dependent $u(1)$ symmetry is associated with $\int d^{d-1}x \, \epsilon \, \mathcal{G} \approx 0$, which involves spacelike derivatives and clearly needs to be supplemented by a non-vanishing surface term in order to be a well-defined canonical generator when the gauge parameter $\epsilon$ does not go to zero.  This surface term can be computed along the lines of \cite{Regge:1974zd}. This is the first global angle-dependent $u(1)$ symmetry. 

It is traditionnally assumed that the generator $\int d^{d-1}x \, \dot{\epsilon} \, \pi^0$, being purely algebraic, does not need to be supplemented by a surface term, no matter how its parameter $ \dot{\epsilon} $ behaves at infinity. This would be true with the standard bulk symplectic form, but is incorrect in the present case due to the surface contribution to the symplectic structure.  As we have seen, the generator $\int d^{d-1}x \, \dot{\epsilon} \, \pi^0$ must be supplemented by a non-vanishing surface term when $ \dot{\epsilon} $ goes to a non-vanishing function of the angles at infinity (at order $\frac1r$). This is the origin of the second angle-dependent global $u(1)$.  

Each first class constraint kills locally one canonical gauge pair and also brings at infinity its own angle-dependent global $u(1)$.  There is therefore symmetry between longitudinal and temporal directions.  

Another way to understand the necessity of the second $u(1)$ comes from Lorentz invariance, once the first $u(1)$ generated by $G[\epsilon]$ is uncovered.  Indeed, we have seen that while $G[\epsilon_0] =Q$ (zero mode = electric charge) is in the trivial representation of the Poincar\'e group, generators $G[\epsilon]$ with a non-trivial angle-dependent  parameter $\epsilon$ do not form by themselves a representation.  They transform into the generator $G[\mu]$ of the second $u(1)$, so that $G[\mu]$ is needed by Lorentz invariance once  $G[\epsilon]$ with a non-trivial angle-dependent  parameter $\epsilon$ appears.   As we pointed out, one has in fact two representations pairing $G[\epsilon]$ and $G[\mu]$ of opposite parities, namely, $(G[\epsilon^{\rm{even}}], G[\mu^{\rm{odd}})$ and $(G[\epsilon^{\rm{odd}}], G[\mu^{\rm{even}})$.  From the point of view of Lorentz invariance, the two angle-dependent $u(1)$'s are more conveniently split in this way.

In four dimensions, parity conditions eliminate one $u(1)$, the one generated by $(G[\epsilon^{\rm{odd}}], G[\mu^{\rm{even}})$.  At the same time, there is no need for imposing 
the condition  $\xbar F_{AB}=0$.  This enables one to consider magnetic sources with monopole decay at infinity.  It has been argued in  \cite{Strominger:2015bla} that a second magnetic $u(1)$ was underlying some soft theorems, so there would be also two $u(1)$'s in that case.

It would be of interest to display explicitly the action of the second $u(1)$ at null infinity and study its quantum implications.  Similarly, the generalization to gravity should be carried out. This will be done in a forthcoming paper \cite{HTToAppear}, where we shall discuss how features similar to those encountered here (absence of parity conditions and expected corresponding enlargement of the asymptotic symmetry - in that case doubling of the BMS supertranslations) should appear in the gravitational context.

\section*{Note added}
While this work was being completed, we received the interesting preprint \cite{Esmaeili:2019hom} that deals with similar questions.  This work confirms the results of \cite{Henneaux:2018gfi}, \cite{Henneaux:2018hdj} that the parity conditions in $d=4$ spacetime dimensions not only make the symplectic form finite but also eliminate the divergence of the fields as one goes to null infinity.  The preprint \cite{Esmaeili:2019hom} also imposes conditions in $d>4$ that are stronger than the conditions discussed here, which explains why only one angle-dependent $u(1)$ is found in that work. Similarly, the recent preprint \cite{He:2019jjk}  considers matching conditions equivalent to the standard parity conditions of $d=4$ and also finds a single $u(1)$.

\section*{Acknowledgements}
We thank Prahar Mitra for conversations that contributed to initiate this work.
This work was partially supported by the ERC Advanced Grant ``High-Spin-Grav'' and by FNRS-Belgium (convention IISN 4.4503.15).

\begin{appendix}

\section{Magnetic Sources in $d>4$}
\label{app:MagnSources}

Let us assume that there are only have magnetic sources.   The equations for the electromagnetic field  takes
the form
\begin{equation}
    \label{eq:magnetEOM}
    \d_\mu F^{\mu\nu} = 0, \qquad
    \d_\mu( \star F)^{\mu \nu_1 \ldots \nu_n} = \kappa^{\nu_1 \ldots
    \nu_n}, \qquad n = d-3,
\end{equation}
where we have assumed usual euclidean coordinates $x^\mu = t, x^1, ...,
x^{d-1}$. We will assume that the magnetic source $\kappa^{\nu_1 \ldots
\nu_n}$ is non-zero in a finite region only. The natural source will be a
$(d-3)$-dimensional object:
\begin{gather}
    Y^\nu(\sigma): \quad \RR \times \Sigma \rightarrow \RR^d, \\
    \kappa^{\nu_1 \ldots \nu_n} (x^\mu) = q \int dY^{\nu_1} \wedge \ldots
    \wedge dY^{\nu_n} \, \delta^d \Big(x^\mu-Y^\mu(\sigma)\Big),
\end{gather}
where $\Sigma$ is a closed spatial manifold of dimension $(d-4)$.

A static configuration will take the following form
\begin{gather}
    Y^0 = \sigma^0, \qquad \d_{\sigma^0} Y^i = 0,\\
    \kappa^{i_1 \ldots i_n} = 0, \quad
    \d_i\kappa^{t i j \ldots j_{n-2}} = 0, \quad
    \kappa^{t i_1 \ldots i_{n-1}}(x^j) = q \oint_\Sigma dY^{i_1} \wedge \ldots
    \wedge dY^{i_{n-1}} \delta^{d-1} \Big( x^j - Y^j(\sigma)\Big), 
\end{gather}
where the time dependence drops out as we can evaluate the integral over
$\sigma^0$. We can compute the "total" charge:
\begin{equation}
    \int d^{d-1}x \, \kappa^{t i_1 \ldots i_{n-1}} = q \oint_\Sigma dY^{i_1} \wedge \ldots
    \wedge dY^{i_{n-1}} = 0.
\end{equation}
In this case, the dynamical equations \eqref{eq:magnetEOM} are easily
solved in terms of a dual potential $B_{\nu_1 \ldots \nu_n}$:
\begin{equation}
    \star F_{\mu \nu_1  \ldots \nu_n} = (n+1) \d_{[\mu} B_{\nu_1 \ldots
    \nu_n]}, \quad
    B_{i_1 \ldots i_n} = 0, \quad
    B^{t i_1 \ldots i_{n-1}} = \Delta^{-1} \kappa^{t i_1 \ldots i_{n-1}}.
\end{equation}
Using the relevant Green's function, we get:
\begin{flalign}
    B^{t i_1 \ldots i_{n-1}}(t,x) & = \int d^{d-1} x' \frac{1}{\vert x - x'
    \vert^{d-3}}  \kappa^{t i_1 \ldots i_{n-1}}(x')\\
    & = \frac{1}{r^{d-3}} \int d^{d-1} x' \kappa^{t i_1 \ldots
    i_{n-1}}(x') + O(r^{2-d})\\
    & =  O(r^{2-d}).
\end{flalign}
The corresponding field strength is given by
\begin{equation}
    \star F^{t i_1 \ldots i_n} = O(r^{1-d}), \quad
    \star F^{i_1 \ldots i_{n+1}} = 0, \quad
    F_{ti} = O(r^{1-d}), \quad F_{ij} = 0.
\end{equation}

\section{Special functions}
\label{app:Special}

\subsection{Spherical harmonics}
\label{sub:SpheHar}

Functions on the $(n-1)$-sphere can be decomposed in spherical harmonics
$Y_l(x^A)$ where $l$ is the degree of the corresponding polynomial\cite{Stein2016}. Spherical harmonics of degree $l$ are
eigenfunctions of the Laplacian on the sphere
\begin{equation}
	\Delta Y_l = - l (n+l-2) Y_l,
\end{equation}
and they have the following parity properties under the antipodal map $x^A
\rightarrow -x^A$
\begin{equation}
	Y_l(-x^A) = (-)^l Y_l (x^A).
\end{equation}

Vectors on the $(n-1)$-sphere can also be decomposed in spherical
harmonics. We have two families: 
\begin{itemize}
\item longitudinal vector fields are decomposed in longitudinal vector
    spherical harmonics $\Phi^A_{l}$. They are eigenfunctions of the
        Laplacian with eigenvalues given by:
        \begin{equation}
            \Delta \Phi^A_{l} = - [l (l+d-2) - d + 2] \Phi^A_{l}, \qquad
            \Phi_{lA} = \d_A Y_{l}
        \end{equation}
\item transverse vector fields are decomposed in transverse vector
    spherical harmonics $\Psi^A_{l}$. They are eigenfunctions of the
        Laplacian with eigenvalues given by:
        \begin{equation}
            \Delta \Psi^A_{l} = - [l (l+d-2) - 1] \Psi^A_l.
        \end{equation}
\end{itemize}

The above results for the decomposition of transverse vector fields lead
to a set of spherical harmonics for exact two forms $\Theta_{lAB} = \d_A
\Psi_{lB} - \d_B \Psi_{lA}$. The action of the Laplacian is given by
\begin{equation}
    \Delta \Theta_{lAB} = - [l (l+d-2) - d+3] \Theta_{lAB}.
\end{equation}

\subsection{Ultraspherical polynomials and functions of the second kind}
\label{sub:Ultra}

This subsection follows very closely \cite{Henneaux:2018mgn}.  We have chosen to give this information again here (rather than referring to the relevant equations in \cite{Henneaux:2018mgn}) for the sake of completeness and for the convenience of the reader.

The central equation for our analysis is (\ref{eq:UltraKey}), i.e., 
\begin{gather}
	\label{eq:alternJacobieq}
    (1-s^2) \d_s^2 Y^{(\lambda)}_n + (2\lambda -3) s \d_s Y^{(\lambda)}_n + (n+1)(n+2\lambda -1)
    Y^{(\lambda)}_n =
	0,\\
    \lambda = k + \frac {d-3}{2}, \quad n = l-k.
\end{gather}
From its definition, the parameter $\lambda$ is an integer (odd spacetime dimensions) or a half-integer (even spacetime dimensions).  Furthermore $\lambda \geq \frac12$, the miminum value $\lambda = \frac12$ being achieved for the leading order of the $\frac{1}{\eta}$ expansion in $4$ spacetime dimensions ($d=4$ and $k=0$).  This minimum value has a special status as we shall see (appeareance of logarithmic singularities).

The equation (\ref{eq:alternJacobieq}) can be brought into  standard form \cite{Ultra} by appropriate changes of variables.  The procedure is different according to whether the parameter $n$, which is an integer, is positive or zero, $n \ge 0$,  or is negative, $n <0$.  Both cases occur in our analysis since $n=l-k$ so that $n <0$ whenever the order $k$ in the $\frac{1}{\eta}$-expansion exceeds the order $l$ of the spherical harmonics.  For the leading $\frac{1}{\eta}$-order $k=0$,  only the case $n \ge 0$ occurs.

\begin{itemize}
    \item $n\ge 0$.
    
The 
rescaling
\begin{equation}
    Y^{(\lambda)}_n(s)  = (1-s^2)^{\lambda-\frac 1 2 } \psi^{(\lambda)}_n(s) \label{eq:Rescaling0}
\end{equation}
brings the equation to the form
\begin{equation}
	\label{eq:diffJacobi}
	(1-s^2) \d_s^2 \psi_n  - (2\lambda +1 ) s \d_s
	\psi_n + n (n+2 \lambda) \psi_n = 0.
\end{equation}
In our case ($\lambda = \frac12, 1, \frac32, \cdots$ and $n \in \NN$), this equation has a polynomial solution. For $\lambda = \frac 1 2$, the equation reduces to the Legendre equation and we recover the familiar Legendre polynomials.
For general $\lambda > - \frac12$, this equation takes a form analysed e.g. in \cite{Ultra}, and the polynomial
solution is known as ultraspherical polynomial or
Gegenbauer's polynomial $P^{(\lambda)}_n$.
These polynomials satisfy
\begin{equation}
	P_n^{(\lambda)}(-s) = (-)^n P_n^{(\lambda)}(s), \qquad P^{(\lambda)}_n(1) = \left
	(\begin{array}{c}n+2\lambda -1 \\ n \end{array}
	\right).
\end{equation}
and can be constructed using the following recurrence
formula
\begin{gather}
	nP^{(\lambda)}_{n}(s) = 2 (n+\lambda-1)s P^{(\lambda)}_{n-1}(s) -
	(n+2\lambda -2) P^{(\lambda)}_{n-2}(s), \qquad n>1,\\
	P^{(\lambda)}_0(s) = 1, \qquad P^{(\lambda)}_1(s) = 2\lambda s.
\end{gather}

The function of the second kind $Q^{(\lambda)}_n$ is the solution of the
differential equation \eqref{eq:diffJacobi} which is linearly independent of
$P^{(\lambda)}_n$. The full set can be constructed using the same recurrence
relation with a different starting point:
\begin{gather}
	nQ^{(\lambda)}_{n}(s) = 2 (n+\lambda-1)s Q^{(\lambda)}_{n-1}(s) -
	(n+2\lambda -2) Q^{(\lambda)}_{n-2}(s), \qquad n>1,\\
	Q^{(\lambda)}_0(s) = \int_0^s(1-x^2)^{-\lambda-\frac 1 2} dx, 
	\qquad Q^{(\lambda)}_1(s) = 2 \lambda s Q^{(\lambda)}_0(s) -
	(1-s^2)^{-\lambda + \frac 1 2 }.
\end{gather}
They take the general form
\begin{equation}
	Q^{(\lambda)}_n(s) = P^{(\lambda)}_n(s) Q^{(\lambda)}_0(s) + 
	R^{(\lambda)}_n(s)(1-s^2)^{-\lambda + \frac 1 2 },
\end{equation}
where $R^{(\lambda)}_n$ are polynomials of degree $n-1$ and satisfy
$Q^{(\lambda)}_n(-s) = (-)^{n+1} Q_n^{(\lambda)}(s)$.
For the values of $\lambda$ relevant for our analysis (half integers and
integers), the functions of the second kind $Q^{(\lambda)}_n(s)$ diverge at $s=\pm 1$. 
For $\lambda = \frac 1 2 $, the Legendre function of the second kind diverges logarithmically while the other values of
$\lambda$ lead to
\begin{equation}
	\lim_{s\to 1} (1-s^2)^{\lambda - \frac 1 2}Q_n^{(\lambda)}(s) = \frac
	1{2\lambda - 1}, \qquad
	n=0,1,\ldots
\end{equation}
($Q_n^{(\lambda)}(s) \sim (1-s^2)^{-\lambda + \frac 1 2}$).

The general solution for $Y^{(\lambda)}_n$ is then given from (\ref{eq:Rescaling0}) by
\begin{gather}
    Y^{(\lambda)}_n(s) =
	A \tilde P^{(\lambda)}_n(s) + B
    \tilde Q^{(\lambda)}_n(s), \\
\tilde P^{(\lambda)}_n(s) = (1-s^2)^{\lambda-\frac 1 2
    }P^{(\lambda)}_n(s), \qquad \tilde Q^{(\lambda)}_n(s) =
    (1-s^2)^{\lambda-\frac 1 2 }Q^{(\lambda)}_n(s), \quad \forall n\ge 0,\\
    \tilde P^{(\lambda)}_n(-s) = (-)^n \tilde P^{(\lambda)}_n(s), \qquad
    \tilde Q^{(\lambda)}_n(-s) = (-)^{n+1} \tilde Q^{(\lambda)}_n(s).
\end{gather}
For all values of $\lambda$, the $\tilde Q$ branch of $Y$ will dominate in the limit
$s\to\pm 1$. If $\lambda = \frac 1 2$, the $\tilde Q$ branch will diverge
logarithmically while the $\tilde P$ branch will be finite. For all other 
values of $\lambda$ (integers and half-integers  $\ge 1$), the $\tilde Q$ branch will be finite and will tend to a
non-zero constant at $s=\pm 1$ while the $\tilde P$ branch will go to zero.

\item $n<0$.

In that case, taking into account that $k \geq 1$ (in order to have $ n = l - k <0$), one finds that $n + 2 \lambda - 1 = l + k + d-4 >0$, so that $n > 1 - 2 \lambda$.  Furthermore $\lambda \geq \frac32 >1$,
$$ 0 > n > 1 - 2 \lambda, \quad \lambda \geq \frac32 .$$

The change of parameters $n= -r -1$ ($r = -n - 1 = 0, 1, 2, \cdots$) and $\lambda = 1 - \rho$ brings the equation to the form 
\begin{gather}
	\label{eq:alternJacobieqBis}
    (1-s^2) \d_s^2 Y^{(\lambda)}_n - (2\rho +1) s \d_s Y^{(\lambda)}_n + r(r+2\rho)
    Y^{(\lambda)}_n =
	0,\\
   r = -n - 1 = 0, 1, 2, \cdots, \quad  \rho = 1- \lambda = 1 - ( k + \frac {d-3}{2}).
\end{gather}
This is again the equation (\ref{eq:diffJacobi}) for ultraspherical polynomials, but with a range of $\rho$ which is not the usual one since $\rho \leq -\frac12$.  This more general equation has been also studied in \cite{Ultra}, where it was found that the pattern is similar: there is a polynomial
branch and a "second" class branch, which is actually also polynomial when $\rho$ (and thus $\lambda$) is a half integer. 

We give the solutions directly in terms of the original parameters $\lambda$ and $n$ appearing in the expansion of the fields.  The polynomial branch $p_r(s)$ of solutions has the standard parity $p_r(-s) = (-)^r p_r(s)$, while the other branch (which might also be polynomial as we pointed out) fulfills $q_r(-s) = (-)^{r+1} q_r(s)$.    Therefore, in order to keep uniformity in the parity properties when expressed in terms of $n$, we denote the polynomial branch by $ \tilde Q^{(\lambda)}_{n}$ and the other branch by $ \tilde P^{(\lambda)}_{n}$. 

The two sets of solutions can be
constructed with the following recurrence relation:
\begin{gather}
    (2\lambda + n -1) \tilde P^{(\lambda)}_{n-1}(s) = 2 (n+\lambda) s
    \tilde P^{(\lambda)}_n(s) - (n+1)
    \tilde P^{(\lambda)}_{n+1}(s),\qquad  \quad n \leq -2, \\
    \tilde P^{(\lambda)}_{-1}(s) =\int_0^s (1-x^2)^{\lambda -
    \frac 3 2} dx, \quad \tilde P^{(\lambda)}_{-2}(s) = s \tilde
    P^{(\lambda)}_{-1}(s) +  \frac 1 {2 (\lambda -1)} (1-s^2)^{\lambda - \frac 1 2 },
\end{gather}
the polynomial branch being given by the following starting point
\begin{equation}
    \tilde Q^{(\lambda)}_{-1}(s) = 1, \quad \tilde Q^{(\lambda)}_{-2}(s) =
    s.
\end{equation}
In order to prove this recurrence, one needs the following relation
\begin{equation}
    (1-s^2) \d_s \tilde P^{(\lambda)}_n = (n+1) (s \tilde P^{(\lambda)}_n
    -\tilde P^{(\lambda)}_{n+1})
\end{equation}
and its equivalent in terms of $\tilde Q$. The parity conditions read
\begin{equation}
    \tilde P^{(\lambda)}_n(-s) = (-)^n \tilde P^{(\lambda)}_n(s), \qquad
    \tilde Q^{(\lambda)}_n(-s) = (-)^{n+1} \tilde Q^{(\lambda)}_n(s).
\end{equation}

The general solution for $Y^{(\lambda)}_n$ keeps the form
\begin{equation}
    Y^{(\lambda)}_n(s) =
	A \tilde P^{(\lambda)}_n(s) + B
    \tilde Q^{(\lambda)}_n(s), \qquad  0>n>1-2\lambda.
\end{equation}
An important difference from the regime $n\ge 0$ is that both branches
have now the same asymptotic behaviour in the limit $s\to \pm 1$: they
both tend to a non-zero finite value. In particular, as we already mentioned, 
both branches are polynomials if $\lambda$ is a half integer.
\end{itemize}

\section{Extra components in hyperbolic description}
\label{app:solEOM}

In the main text, we have solved  the equations of motion for one
component of the curvature $F_{\mu\nu}$, namely, $E^s$. In this appendix, we will solve
for the other components.

The equations to be solved are
\begin{gather}
    E_b = F_{b\eta} = \sum_{k\geq 0} \eta^{-d + 3 - k} E_b^{(k)}, \qquad
    F_{ab} = \sum_{k \geq 0} \eta^{-d + 4 - k} F_{ab}^{(k)}, \qquad
    F_{ab}^{(0)} = 0,\\
    \label{eq:maxI}
    \mathcal D^a E^{(k)}_a = 0, \qquad (d-4+k) F_{ab}^{(k)} = \d_a E^{(k)}_b - \d_b E^{(k)}_a,
\\
    \label{eq:maxII}
     \mathcal D^b F^{(k)}_{ba} + k E^{(k)}_a = 0,    \qquad
    \d_{[a} F^{(k)}_{bc]} = 0.
\end{gather}
Combining these equations, we can obtain second order equations for both
$E_a^{(k)}$ and $F_{ab}^{(k)}$:
\begin{flalign}
    \label{eq:EOMappI}
    \mathcal D^a \mathcal D_a E_b^{(k)} -(d-2) E_b^{(k)} + k(d-4 + k)E_b^{(k)}& = 0,\\
    \label{eq:EOMappII}
    \mathcal D^a \mathcal D_a F^{(k)}_{bc} - 2(d-3) F^{(k)}_{bc} +
    k(d-4+k) F^{(k)}_{bc} & = 0.
\end{flalign}
As we saw in the main text, the first equation combined with the fact that
$E_a^{(k)}$ is divergence-less on the hyperboloid implies an evolution
equation for $E_s^{(k)}$:
\begin{equation}
    \label{eq:evolEs}
    (1-s^2) \d_s^2 E^{(k)}_s + (d-6)s \d_s  E^{(k)}_s + (d-4) E^{(k)}_s -
    D_A D^A E^{(k)}_s  - (1-s^2)^{-1}k (d-4+k) E^{(k)}_s=0.
\end{equation}
In a similar way, the second equation combined with the fact that
$F_{ab}^{(k)}$ is closed on the hyperboloid gives us
an evolution equation for $F_{AB}^{(k)}$:
\begin{multline}
    \label{eq:evolFAB}
    (1-s^2) \d_s^2 F^{(k)}_{AB} + (d-6) s \d_s  F^{(k)}_{AB}+ 2 (d-4)  F^{(k)}_{AB}
    \\- D^C D_C  F^{(k)}_{AB} - (1-s^2)^{-1} k(d-4+k)  F^{(k)}_{AB}= 0.
\end{multline}

The general solution for $E_s^{(k)}$ is given
in terms of ultraspherical polynomials as follows
\begin{equation}
    E_s^{(k)} = (1-s^2)^{-\frac k 2} \sum_{l,m}\Big( \Xi_{lm}^{P(k)} \tilde
    P_n^{(\lambda)} + \Xi_{lm}^{Q(k)} \tilde
    Q_n^{(\lambda)}\Big) Y_{lm},
\end{equation}
where $\lambda = k + \frac {d-3} 2$ and $n=l-k$. In order to obtain the
general solution to equation \eqref{eq:evolFAB}, we will use the same
strategy and decompose
$F_{AB}^{(k)}$ in spherical harmonics:
\begin{equation}
    F_{AB}^{(k)} = (1-s^2)^{-\frac k 2} \sum_{l>0,m} \, \alpha^{(k)}_{lm}
    \Theta_{lmAB}.
\end{equation}
The exact spherical harmonics $\Theta_{lmAB}$ are enough as $F_{AB}^{(k)}$
is  an exact 2-form on the sphere. Equation \eqref{eq:evolFAB} then takes the form
\begin{equation}
    (1-s^2) \d_s^2 \alpha_{lm}^{(k)} + (d+2k-6) s \d_s  \alpha_{lm}^{(k)}+ 
    l(l+d-3) \alpha_{lm}^{(k)} -(k-1)(k + d-4) \alpha_{lm}^{(k)}= 0,
\end{equation}
which has the following solution
\begin{equation}
    \alpha_{lm}^{(k)} = \alpha_{lm}^{P(k)}
    \tilde P^{(\lambda)}_n (s) + \alpha_{lm}^{Q(k)}
    \tilde Q^{(\lambda)}_n (s),
\end{equation}
with the same $\lambda$ and $n$.

Using the expression for $E_s^{(k)}$ and $F_{AB}^{(k)}$, we
can now build the remaining components of the curvature in such a way that the
main equations are satisfied. This is done by expending 
\eqref{eq:maxI}: 
\begin{flalign}
    \label{eq:EAL}
    D^A E^{(k)}_A  & = (1-s^2)\d_s E^{(k)}_s + (d-4) s E_s^{(k)}, \\
    \label{eq:EAT}
    \d_A E^{(k)}_B - \d_B E^{(k)}_A & = (d-4+k) F^{(k)}_{AB}, \\
    \label{eq:FsA}
    F^{(k)}_{sA} & = \frac 1 {d-4+k} (\d_s E^{(k)}_A - \d_A E^{(k)}_s).
\end{flalign}
The first two lines give respectively the longitudinal and the transverse
part of $E_A^{(k)}$ in terms of $E_s^{(k)}$ and $F_{AB}^{(k)}$ while the
last one then gives the last components $F_{sA}^{(k)}$ in term of the
rest. The only consistency condition coming from these equations concerns
the zero-mode of \eqref{eq:EAL}. Integrating this equation on the
sphere, we see that the zero-mode of the RHS has to be zero:
\begin{multline}
    0 = \int d^{d-2}\Omega \, \Big\{(1-s^2)\d_s E^{(k)}_s + (d-4) s
    E_s^{(k)}\Big\}\\ = (1-s^2)^{-\frac k 2}
    \Big((1-s^2) \d_s  + (d-4+k) s\Big)\Big(\Xi_{00}^{P(k)} \tilde
    P^{(\lambda)}_{-k} + \Xi_{00}^{Q(k)}\tilde Q^{(\lambda)}_{-k}\Big).
\end{multline}
This implies that $\Xi_{00}^{P(k)} \tilde P^{(\lambda)}_{-k} +
\Xi_{00}^{Q(k)}\tilde Q^{(\lambda)}_{-k}$ has to be proportional to
$(1-s^2)^{\frac {d-4+k}2}$. One can easily check that, except for $k=0$
where $P^{(\lambda)}_{0} = (1-s^2)^{\frac {d-4}2}$, the function
$(1-s^2)^{\frac {d-4+k}2}$ is not a solution of the dynamical equation of
the zero-mode of $\Xi^{(k)}$. This implies
\begin{equation}
    \Xi_{00}^{P(k)} = \Xi_{00}^{P(0)}\delta_0^k, \qquad \Xi_{00}^{Q(k)} = 0.
\end{equation}

Using the symbols $T$ and $L$ to respectively denote the transverse
and longitudinal part of vectors defined on the sphere, we obtain the
following expressions
\begin{flalign}
    E^{(k)T}_A &= (d-4+k) (1-s^2)^{-\frac k 2} \sum_{l>0,m} \Big(\alpha_{lm}^{P(k)}
    \tilde P^{(\lambda)}_n + \alpha_{lm}^{Q(k)}
    \tilde Q^{(\lambda)}_n\Big)
    \Psi_{lmA}, \\
    E^{(k)L}_A &= -  (1-s^2)^{-\frac k 2} \sum_{l>0,m}
    \frac 1 {l (l+d-3)}\nonumber\\ & \qquad \qquad \qquad\Big((1-s^2)\d_s +
    (d-4+k)s\Big)\Big( \Xi_{lm}^{P(k)} \tilde
    P_n^{(\lambda)} + \Xi_{lm}^{Q(k)} \tilde
    Q_n^{(\lambda)}\Big)\Phi_{lmA},\\
    F^{(k)T}_{sA} &=   (1-s^2)^{-\frac k 2-1}
    \sum_{l>0,m}\Big((1-s^2) \d_s + ks\Big) \Big(\alpha_{lm}^{P(k)}
    \tilde P^{(\lambda)}_n + \alpha_{lm}^{Q(k)}
    \tilde Q^{(\lambda)}_n\Big)
    \Psi_{lmA}, \\
    F^{(k)L}_{sA} &= -  k (1-s^2)^{-\frac k 2-1} \sum_{l>0,m}
    \frac 1 {l (l+d-3)}\Big( \Xi_{lm}^{P(k)} \tilde
    P_n^{(\lambda)} + \Xi_{lm}^{Q(k)} \tilde
    Q_n^{(\lambda)}\Big)\Phi_{lmA}.
\end{flalign}
This provides the general solution to the complete system as the tensors
$F_{ab}^{(k)}$ and $E_a^{(k)}$ built in this way satisfy the last two
equations given in \eqref{eq:maxII}. The second equation is trivially
satisfied while the first one can be shown to reduce to combinations of
the two evolution equations \eqref{eq:evolEs} and \eqref{eq:evolFAB}.

\end{appendix}


\begin{thebibliography}{99}

\bibitem{Bondi:1962px}
  H.~Bondi, M.~G.~J.~van der Burg and A.~W.~K.~Metzner,
  ``Gravitational waves in general relativity. 7. Waves from axisymmetric isolated systems,''
  Proc.\ Roy.\ Soc.\ Lond.\ A {\bf 269} (1962) 21.
  
\bibitem{Sachs:1962wk}
  R.~K.~Sachs,
  ``Gravitational waves in general relativity. 8. Waves in asymptotically flat space-times,''
  Proc.\ Roy.\ Soc.\ Lond.\ A {\bf 270} (1962) 103.
  
\bibitem{Sachs:1962zza}
  R.~Sachs,
  ``Asymptotic symmetries in gravitational theory,''
  Phys.\ Rev.\  {\bf 128} (1962) 2851.

\bibitem{Penrose:1962ij}
  R.~Penrose,
  ``Asymptotic properties of fields and space-times,''
  Phys.\ Rev.\ Lett.\  {\bf 10} (1963) 66.
  
\bibitem{Madler:2016xju}
  T.~M\"adler and J.~Winicour,
  ``Bondi-Sachs Formalism,''
  Scholarpedia {\bf 11} (2016) 33528
  [arXiv:1609.01731 [gr-qc]]. (Review)
  
\bibitem{Alessio:2017lps}
  F.~Alessio and G.~Esposito,
  ``On the structure and applications of the Bondi-Metzner-Sachs group,''
  arXiv:1709.05134 [gr-qc]. (Review)
  
   \bibitem{Ashtekar:2018lor}
  A.~Ashtekar, M.~Campiglia and A.~Laddha,
  ``Null infinity, the BMS group and infrared issues,''
  Gen.\ Rel.\ Grav.\  {\bf 50} (2018) no.11,  140
  [arXiv:1808.07093 [gr-qc]].


\bibitem{Christodoulou:1993uv}
  D.~Christodoulou and S.~Klainerman,
  ``The Global nonlinear stability of the Minkowski space,''
  Princeton University Press, Princeton, 1993

\bibitem{Friedrich:2017cjg}
  H.~Friedrich,
  ``Peeling or not peeling - is that the question?,''
  Class.\ Quant.\ Grav.\  {\bf 35} (2018) no.8,  083001
  [arXiv:1709.07709 [gr-qc]].
  
  
\bibitem{Strominger:2013jfa}
  A.~Strominger,
  ``On BMS Invariance of Gravitational Scattering,''
  JHEP {\bf 1407} (2014) 152
  [arXiv:1312.2229 [hep-th]].
  
\bibitem{He:2014laa}
  T.~He, V.~Lysov, P.~Mitra and A.~Strominger,
  ``BMS supertranslations and Weinberg's soft graviton theorem,''
  JHEP {\bf 1505} (2015) 151
  [arXiv:1401.7026 [hep-th]].
  
\bibitem{Cachazo:2014fwa}
  F.~Cachazo and A.~Strominger,
  ``Evidence for a New Soft Graviton Theorem,''
  arXiv:1404.4091 [hep-th].
  
\bibitem{Strominger:2014pwa}
  A.~Strominger and A.~Zhiboedov,
  ``Gravitational Memory, BMS Supertranslations and Soft Theorems,''
  JHEP {\bf 1601} (2016) 086
  [arXiv:1411.5745 [hep-th]].
  
\bibitem{Pasterski:2015tva}
  S.~Pasterski, A.~Strominger and A.~Zhiboedov,
  ``New Gravitational Memories,''
  JHEP {\bf 1612} (2016) 053
  [arXiv:1502.06120 [hep-th]].

\bibitem{Campiglia:2015kxa}
  M.~Campiglia and A.~Laddha,
  ``Asymptotic symmetries of gravity and soft theorems for massive particles,''
  JHEP {\bf 1512} (2015) 094
  [arXiv:1509.01406 [hep-th]].

\bibitem{Conde:2016rom}
  E.~Conde and P.~Mao,
  ``BMS Supertranslations and Not So Soft Gravitons,''
  JHEP {\bf 1705} (2017) 060
  [arXiv:1612.08294 [hep-th]].

  
\bibitem{Strominger:2017zoo}
  A.~Strominger,
  ``Lectures on the Infrared Structure of Gravity and Gauge Theory,''
  arXiv:1703.05448 [hep-th].
  
\bibitem{Ashtekar:1981bq}
  A.~Ashtekar and M.~Streubel,
  ``Symplectic Geometry of Radiative Modes and Conserved Quantities at Null Infinity,''
  Proc.\ Roy.\ Soc.\ Lond.\ A {\bf 376} (1981) 585.
  
\bibitem{Ashtekar:1981sf}
  A.~Ashtekar,
  ``Asymptotic Quantization of the Gravitational Field,''
  Phys.\ Rev.\ Lett.\  {\bf 46} (1981) 573.
  
\bibitem{Ashtekar:1987tt}
  A.~Ashtekar,
  ``Asymptotic Quantization: Based On 1984 Naples Lectures,''
  Naples, Italy: Bibliopolis (1987) (Monographs and Textbooks in Physical Science, 2)
  
\bibitem{Banks:2003vp}
  T.~Banks,
  ``A Critique of pure string theory: Heterodox opinions of diverse dimensions,''
  hep-th/0306074.
   
\bibitem{Barnich:2010eb}
  G.~Barnich and C.~Troessaert,
  ``Aspects of the BMS/CFT correspondence,''
  JHEP {\bf 1005} (2010) 062
  [arXiv:1001.1541 [hep-th]].
  
\bibitem{Barnich:2009se}
  G.~Barnich and C.~Troessaert,
  ``Symmetries of asymptotically flat 4 dimensional spacetimes at null infinity revisited,''
  Phys.\ Rev.\ Lett.\  {\bf 105} (2010) 111103
    [arXiv:0909.2617 [gr-qc]].

\bibitem{Arnowitt:1962hi}
  R.~L.~Arnowitt, S.~Deser and C.~W.~Misner,
  ``The Dynamics of general relativity,''
  Gen.\ Rel.\ Grav.\  {\bf 40} (2008) 1997
  [gr-qc/0405109].
  
\bibitem{Regge:1974zd}
  T.~Regge and C.~Teitelboim,
  ``Role of Surface Integrals in the Hamiltonian Formulation of General Relativity,''
  Annals Phys.\  {\bf 88} (1974) 286.
  
\bibitem{Geroch:1972up}
  R.~P.~Geroch,
  ``Structure of the gravitational field at spatial infinity,''
  J.\ Math.\ Phys.\  {\bf 13} (1972) 956.
  
\bibitem{Ashtekar:1978zz}
  A.~Ashtekar and R.~O.~Hansen,
  ``A unified treatment of null and spatial infinity in general relativity. I - Universal structure, asymptotic symmetries, and conserved quantities at spatial infinity,''
  J.\ Math.\ Phys.\  {\bf 19} (1978) 1542.
  
\bibitem{Ashtekar:1991vb}
  A.~Ashtekar and J.~D.~Romano,
  ``Spatial infinity as a boundary of space-time,''
  Class.\ Quant.\ Grav.\  {\bf 9} (1992) 1069.
  
\bibitem{Henneaux:2018hdj}
  M.~Henneaux and C.~Troessaert,
  ``Hamiltonian structure and asymptotic symmetries of the Einstein-Maxwell system at spatial infinity,''
  JHEP {\bf 1807} (2018) 171
  [arXiv:1805.11288 [gr-qc]].
  
\bibitem{Henneaux:2018cst}
  M.~Henneaux and C.~Troessaert,
  ``BMS Group at Spatial Infinity: the Hamiltonian (ADM) approach,''
  JHEP {\bf 1803} (2018) 147
  [arXiv:1801.03718 [gr-qc]].
  
\bibitem{Troessaert:2017jcm}
  C.~Troessaert,
  ``The BMS4 algebra at spatial infinity,''
  Class.\ Quant.\ Grav.\  {\bf 35} (2018) no.7,  074003
   [arXiv:1704.06223 [hep-th]].
 
\bibitem{Strominger:2013lka}
  A.~Strominger,
  ``Asymptotic Symmetries of Yang-Mills Theory,''
  JHEP {\bf 1407} (2014) 151
  [arXiv:1308.0589 [hep-th]].
  
\bibitem{Barnich:2013sxa}
  G.~Barnich and P.~H.~Lambert,
  ``Einstein-Yang-Mills theory: Asymptotic symmetries,''
  Phys.\ Rev.\ D {\bf 88} (2013) 103006
   [arXiv:1310.2698 [hep-th]].

\bibitem{He:2014cra}
  T.~He, P.~Mitra, A.~P.~Porfyriadis and A.~Strominger,
  ``New Symmetries of Massless QED,''
  JHEP {\bf 1410} (2014) 112
  [arXiv:1407.3789 [hep-th]].
  
\bibitem{Henneaux:1999ct}
  M.~Henneaux, B.~Julia and S.~Silva,
  ``Noether superpotentials in supergravities,''
  Nucl.\ Phys.\ B {\bf 563} (1999) 448
  doi:10.1016/S0550-3213(99)00536-2
  [hep-th/9904003].
  
\bibitem{Henneaux:2018gfi}
  M.~Henneaux and C.~Troessaert,
  ``Asymptotic symmetries of electromagnetism at spatial infinity,''
  JHEP {\bf 1805} (2018) 137
  [arXiv:1803.10194 [hep-th]].
    
\bibitem{Tanabe:2012fg}
  K.~Tanabe, T.~Shiromizu and S.~Kinoshita,
  ``Angular momentum at null infinity in higher dimensions,''
  Phys.\ Rev.\ D {\bf 85} (2012) 124058
  [arXiv:1203.0452 [gr-qc]].
  
\bibitem{Kapec:2015vwa}
  D.~Kapec, V.~Lysov, S.~Pasterski and A.~Strominger,
  ``Higher-Dimensional Supertranslations and Weinberg's Soft Graviton Theorem,''
  Annals of Mathematical Sciences and Applications, Volume 2 (2017),
  pp 69-94
  [arXiv:1502.07644 [gr-qc]].
  
\bibitem{Hollands:2016oma}
  S.~Hollands, A.~Ishibashi and R.~M.~Wald,
  ``BMS Supertranslations and Memory in Four and Higher Dimensions,''
  Class.\ Quant.\ Grav.\  {\bf 34} (2017) no.15,  155005
  [arXiv:1612.03290 [gr-qc]].
  
\bibitem{Garfinkle:2017fre}
  D.~Garfinkle, S.~Hollands, A.~Ishibashi, A.~Tolish and R.~M.~Wald,
  ``The Memory Effect for Particle Scattering in Even Spacetime Dimensions,''
  Class.\ Quant.\ Grav.\  {\bf 34} (2017) no.14,  145015
  [arXiv:1702.00095 [gr-qc]].
  
\bibitem{Mao:2017wvx}
  P.~Mao and H.~Ouyang,
  ``Note on soft theorems and memories in even dimensions,''
  Phys.\ Lett.\ B {\bf 774} (2017) 715
  [arXiv:1707.07118 [hep-th]].
  
\bibitem{Campiglia:2017xkp}
  M.~Campiglia and L.~Coito,
  ``Asymptotic charges from soft scalars in even dimensions,''
  Phys.\ Rev.\ D {\bf 97} (2018) no.6,  066009
  [arXiv:1711.05773 [hep-th]].
  
\bibitem{Pate:2017fgt}
  M.~Pate, A.~M.~Raclariu and A.~Strominger,
  ``Gravitational Memory in Higher Dimensions,''
  JHEP {\bf 1806} (2018) 138
  [arXiv:1712.01204 [hep-th]].
  
\bibitem{Campoleoni:2017qot}
  A.~Campoleoni, D.~Francia and C.~Heissenberg,
  ``Asymptotic Charges at Null Infinity in Any Dimension,''
  Universe {\bf 4} (2018) no.3,  47
  [arXiv:1712.09591 [hep-th]].
  
\bibitem{Afshar:2018apx}
  H.~Afshar, E.~Esmaeili and M.~M.~Sheikh-Jabbari,
  ``Asymptotic Symmetries in $p$-Form Theories,''
  JHEP {\bf 1805} (2018) 042
  [arXiv:1801.07752 [hep-th]].
  
\bibitem{Campoleoni:2018uib}
  A.~Campoleoni, D.~Francia and C.~Heissenberg,
  ``Asymptotic symmetries and charges at null infinity: from low to high spins,''
  EPJ Web Conf.\  {\bf 191} (2018) 06011
  [arXiv:1808.01542 [hep-th]].
  
\bibitem{Hollands:2003ie}
  S.~Hollands and A.~Ishibashi,
  ``Asymptotic flatness and Bondi energy in higher dimensional gravity,''
  J.\ Math.\ Phys.\  {\bf 46} (2005) 022503
  [gr-qc/0304054].
  
\bibitem{Hollands:2004ac}
  S.~Hollands and R.~M.~Wald,
  ``Conformal null infinity does not exist for radiating solutions in odd spacetime dimensions,''
  Class.\ Quant.\ Grav.\  {\bf 21} (2004) 5139
  [gr-qc/0407014].
  
\bibitem{Tanabe:2011es}
  K.~Tanabe, S.~Kinoshita and T.~Shiromizu,
  ``Asymptotic flatness at null infinity in arbitrary dimensions,''
  Phys.\ Rev.\ D {\bf 84} (2011) 044055
  [arXiv:1104.0303 [gr-qc]].
  
\bibitem{Campiglia:2017mua}
  M.~Campiglia and R.~Eyheralde,
  ``Asymptotic $U(1)$ charges at spatial infinity,''
  JHEP {\bf 1711} (2017) 168
  [arXiv:1703.07884 [hep-th]].
  
\bibitem{Campiglia:2018dyi}
  M.~Campiglia and A.~Laddha,
  ``Asymptotic charges in massless QED revisited: A view from Spatial Infinity,''
  arXiv:1810.04619 [hep-th].
  
     \bibitem{BeigSchmidt}
  R.~Beig and B.~Schmidt, ``Einstein's equations near spatial infinity,'' Commun. Math. Phys. {\bf 87} (1982) 65.
  
\bibitem{Beig:1983sw}
  R.~Beig,
  ``Integration Of Einstein's Equations Near Spatial Infinity,''
  Proc.  Royal Soc. A
{\bf 1801} (1984) 295--304.
  
\bibitem{Compere:2011ve}
  G.~Compere and F.~Dehouck,
  ``Relaxing the Parity Conditions of Asymptotically Flat Gravity,''
  Class.\ Quant.\ Grav.\  {\bf 28} (2011) 245016
   Erratum: [Class.\ Quant.\ Grav.\  {\bf 30} (2013) 039501]
  [arXiv:1106.4045 [hep-th]].

   \bibitem{Fried1}
  H. Friedrich, ``Gravitational fields near space-like and null infinity,'' J. Geom. Phys. {\bf 24} (1998) 83-163.
  
\bibitem{Friedrich:1999wk}
  H.~Friedrich and J.~Kannar,
  ``Bondi type systems near space - like infinity and the calculation of the NP constants,''
  J.\ Math.\ Phys.\  {\bf 41} (2000) 2195
  [gr-qc/9910077].
  
\bibitem{Friedrich:1999ax}
  H.~Friedrich and J.~Kannar,
  ``Calculating asymptotic quantities near space - like and null infinity from Cauchy data,''
  Annalen Phys.\  {\bf 9} (2000) 321
  [gr-qc/9911103].
  
\bibitem{Benguria:1976in}
  R.~Benguria, P.~Cordero and C.~Teitelboim,
  ``Aspects of the Hamiltonian Dynamics of Interacting Gravitational Gauge and Higgs Fields with Applications to Spherical Symmetry,''
  Nucl.\ Phys.\ B {\bf 122} (1977) 61.
  
  \bibitem{Fock} Vladimir A. Fock, ``The Theory of Space, Time and
Gravitation'', 1st Edition, GITTL, Moscow, 1955. sec 53: 54; 2nd
Revised Edition, Pergamon Press (Oxford: 1964).
  
\bibitem{Deser:2019acl}
  S.~Deser,
  ``Energy in Gravitation and Noether's Theorems,''
  arXiv:1902.05105 [gr-qc].
   
\bibitem{Teitelboim:1985yc} 
  C.~Teitelboim,
  ``Monopoles of Higher Rank,''
  Phys.\ Lett.\  {\bf 167B}, 69 (1986).

\bibitem{Henneaux:1990au}
  M.~Henneaux, C.~Teitelboim and J.~Zanelli,
  ``Gauge Invariance and Degree of Freedom Count,''
  Nucl.\ Phys.\ B {\bf 332} (1990) 169.
  
\bibitem{Henneaux:2018mgn}
  M.~Henneaux and C.~Troessaert,
  ``Asymptotic structure of a massless scalar field and its dual two-form field at spatial infinity,''
  arXiv:1812.07445 [hep-th].
  
\bibitem{Satishchandran:2019pyc}
  G.~Satishchandran and R.~M.~Wald,
  ``The Asymptotic Behavior of Massless Fields and the Memory Effect,''
  arXiv:1901.05942 [gr-qc].
  
\bibitem{Strominger:2015bla}
  A.~Strominger,
  ``Magnetic Corrections to the Soft Photon Theorem,''
  Phys.\ Rev.\ Lett.\  {\bf 116} (2016) no.3,  031602
  [arXiv:1509.00543 [hep-th]].

\bibitem{HTToAppear}
M. Henneaux and C. Troessaert, in preparation.
  
\bibitem{Esmaeili:2019hom}
  E.~Esmaeili,
  ``Asymptotic Symmetries of Maxwell Theory in Arbitrary Dimensions at Spatial Infinity,''
  arXiv:1902.02769 [hep-th].
  
\bibitem{He:2019jjk}
  T.~He and P.~Mitra,
  ``Asymptotic Symmetries and Weinberg's Soft Photon Theorem in Mink$_{d+2}$,''
  arXiv:1903.02608 [hep-th].
  
  
\bibitem{Stein2016}
E.M.~Stein and G.~Weiss, ``Introduction to Fourier analysis on Euclidean spaces'',
PMS volume 32 {2016} (Princeton university press)

    \bibitem{Ultra}
 G. Szeg\H{o}, ``Orthogonal Polynomials'',  Colloquium Publications of the American Mathematical Society, Volume 23, fourth edition (Providence: 1975) (Chapter IV, in particular section 4.7)


\end{thebibliography}
\end{document}